\documentclass[10pt,aps,prd,twocolumn,a4paper,showpacs,nofootinbib]{revtex4-1}
\usepackage[utf8x]{inputenc}
\usepackage{amsmath}
\usepackage{amsfonts}
\usepackage[colorlinks=true,linkcolor=blue,citecolor=blue,urlcolor=blue]{hyperref}
\usepackage{amssymb}
\usepackage{graphicx}
\usepackage{csquotes}

\renewcommand{\Im}{\textrm{Im}}
\newcommand{\bk}{\mathbf{k}}
\newcommand{\eq}[1]{Eq.~\eqref{#1}}
\providecommand{\abs}[1]{\left|#1\right|}
\providecommand{\ep}[1]{{e}^{#1}}

\begin{document}
\title{Dynamical Casimir Effect in dissipative media: \texorpdfstring{\\}{}
 When is the final state non-separable ? }

\author{Xavier Busch}
\email[]{xavier.busch@th.u-psud.fr}
\author{Renaud Parentani}
\email[]{renaud.parentani@th.u-psud.fr}
\affiliation{Laboratoire de Physique Th\'eorique, CNRS UMR 8627, B{\^{a}}timent\ 210,
         \\Universit\'e Paris-Sud 11, 91405 Orsay CEDEX, France}
         
         
\begin{abstract}
We study the consequences of dissipation in homogeneous media when the system is subject to a sudden change, thereby producing pairs of correlated quasi-particles with opposite momenta. We compute both the modifications of the spectrum, and those of the correlations. In particular, we compute the final coherence level, and identify the regimes where the state is non-separable. To isolate the role of dissipation, we first consider dispersive media and study the competition between the intensity of the jump which induces some coherence, and the temperature which reduces it. The contributions of  stimulated and spontaneous emission are clearly identified. We then study how dissipation modifies this competition.
\end{abstract}

\maketitle
\section{Introduction}

The analogy~\cite{Unruh:1980cg} between sound propagation in non-uniform fluids and light propagation in curved space-times opens the possibility to experimentally test long standing predictions of quantum field theory~\cite{birrell1984quantum}, such as the origin of the large scale structures in our Universe~\cite{Mukhanov:1981xt,mukhanov2005physical}, and the Hawking radiation emitted by black holes. In the first case, the cosmic expansion engenders a parametric amplification of homogeneous modes which is very similar to that at the origin of the Dynamical Casimir Effect (DCE)~\cite{PhysRevLett.109.220401,WilsonDCE}, compare for instance \cite{Campo:2003pa} with \cite{Carusotto:2009re}. Yet, in order to predict with accuracy what should be observed, one must take into account the short distance properties of the medium because the predictions involve short wavelength modes. As a result, the analogy breaks down and a case by case analysis is required. 

In homogeneous isotropic media, quasi-particle excitations are governed by dispersion relations of the form
\begin{equation}
\begin{split}
\Omega^2  +  2 i \,  \Omega \, \Gamma(k^2) = c^2 k^2 + f(k^2)~. 
\label{dr}
\end{split}
\end{equation}
In this equation, $\Omega$ is the frequency in the medium frame, $k$ the wave vector, and $c^2$ the low frequency group velocity. The real function $f$ characterizes the elastic (norm preserving) high momentum dispersive effects. The absorptive properties are described by the imaginary contribution governed by $\Gamma> 0$. So far, following~\cite{Unruh:1994je} most works analyzed the consequences of short distance dispersion, see~\cite{Balbinot:2006ua,Barcelo:2005fc,Robertson:2012ku,Jacobson:2012ei} for reviews. Comparatively, much little attention has been paid to dissipative effects. Following~\cite{Parentani:2007uq,Adamek:2008mp,Adamek:2013vw}, in this paper we study the consequences of dissipation when a sudden change is applied to a homogeneous system. 

Because of the homogeneity and the gaussianity of the systems we shall study, the states are composed of two-mode sectors of opposite wave vectors $\{\bk,-\bk\}$ which do not interact with each other. Hence, each sector can be studied separately. Besides the expected temporal decay of the out of equilibrium distribution, we shall see that the mean particle number $n_k = {\rm Tr}[\hat \rho_T \, \hat a_\bk^\dagger \hat a_{\bk}]$ is not significantly affected by turning on dissipation in \eq{dr}. On the contrary, the issue of the correlations between quasi-particles of opposite $\bk$ is more tricky to handle. First, when late time dissipation is sufficiently small that the state can be meaningfully decomposed using a particle number basis, the complex number $c_k = {\rm Tr}[\hat \rho_T \, \hat a_\bk \hat a_{-\bk}]$ accounts for the strength of these correlations. Second, to determine if the final state is quantum mechanically entangled, one should consider the relative value of the norm of $c_k$ with respect to $n_k$~\cite{Campo:2004sz,Campo:2005sy,Campo:05sv20}. Indeed, whenever $\Delta_k$ given by 
\begin{equation}
\begin{split}
\Delta_k \doteq n_k - \abs{c_k}~, 
\label{deltak}
\end{split}
\end{equation}
is negative, the bi-partite state $\{\bk,-\bk\}$ is non separable, i.e., so correlated that its statistical properties cannot be reproduced by a stochastic ensemble. Other means can also be envisaged, such as Cauchy-Schwarz inequalities~\cite{PhysRevLett.108.260401,deNova:2012hm,Adamek:2013vw}, or sub-Poissonian statistics~\cite{mandel1995optical}. Third, as we shall see, even in the absence of dissipation, it is difficult to produce non-separable states, as an initial temperature increases the value of $\Delta_k$ (because it increases the contribution of stimulated emission). The difficulties are reinforced in the presence of dissipation. Indeed, the coupling to the environment generally induces an increase of $\Delta_k$. Our  principal aim is to study the outcome of the competition between the sudden change, which produces the coherence, and the combined effect of temperature and dissipation which reduce it. 

The paper is organized as follows. In Section~\ref{sec:action}, we explain how to couple the phonon field to an environment so as to engender some specific decay rate. In Section~\ref{sec:nodiss}, we study the effect of the temperature on the final value of $\Delta_k$ in the case there is no dissipation. In Section~\ref{sec:dissipative} we study the modifications on the spectrum and $\Delta_k$ when including dissipation. We conclude in Section~\ref{sec:conclu}.

\section{Actions for dissipative phonons}
\label{sec:action}

To study the phenomenology of the DCE in dissipative media, we work with an action of the form $S_T = S_\phi + S_\Psi + S_{\rm int}$, where $\hat \phi$ describes the (free) quasi-particles, $\hat \Psi$ describes the (environmental) degrees of freedom that shall cause the dissipation, and $S_{\rm int}$ describes the interactions between $\hat \phi$ and $\hat \Psi$. As usually done in atomic damping~\cite{Aichelburg1976264} or when describing quantum Brownian motion~\cite{Unruh:1989dd,gardiner2004quantum}, the action $S_T$ is taken quadratic in $\hat \phi,\hat \Psi$, so that the field equations are linear. 

In experiments, a finite range of $k$ shall be accessible. The strength of the correlations can thus be studied as a function of $k$. To cover various cases, we consider decay rates parametrized by 
\begin{equation}
\begin{split}
\Gamma(k) = g^2 \, {(c / \xi)} \,   (\rho \xi)^{2\alpha - 1} {(k \xi)^{2+2 n}} ~. 
\label{Gammapheno}
\end{split}
\end{equation}
The coupling constant $ g$ is dimensionless, $\rho$ is the condensed atoms density, and $\xi$ a short distance length which corresponds to the healing length in atomic Bose gases. To fix the ideas and notations, the quasi particles shall be phonons propagating in an elongated, effectively one-dimensional, atomic Bose condensate, as in the experiment of~\cite{PhysRevLett.109.220401}. However our treatment can be easily adapted to other systems displaying dissipation such as polariton excitations in micro-cavities~\cite{Gerace:2012an}.\footnote{
While finalizing this work, we became aware of \cite{Koghee2013} where similar issues are considered in that context. Let us signal several differences. First, in the present work, dissipation is handled in a way that allows to compute correlation functions at different times. Second, we keep the contribution of $c_k = {\rm Tr}[\hat \rho_T \, \hat a_\bk \hat a_{-\bk}]$ which accounts for the correlations, and use it to distinguish the states that are non-separable. 
Third, our dissipative effects only affect the quasi-particles, see Appendix~\ref{app:condensation}. } 
The powers $n$ and $\alpha$ can be chosen to reproduce effects computed from first principles. For instance, in Bose gases, two types of dissipative effects are found: The first one, called Beliaev decay~\cite{pitaevskii2003bose}, scales with $\alpha=0$ and $n=3/2$, while the second one, the Landau decay~\cite{pethick2002bose}, depends on the temperature, has also $\alpha=0$, and is proportional to $ck \sqrt{1 + k^2 \xi^2}$.

In this paper, $\hat \phi$ describes relative density fluctuations propagating in homogeneous time dependent condensates, see Appendix \ref{app:condensation} for details. Working with $\hbar= 1$, its action is
\begin{equation}
\label{SPhi}
\begin{split}
S_\phi = \! \frac12 \int \! \! dt dx \rho \big \{ & \hat \phi^\dagger [i \partial_t + \frac{\partial_x^2 }{2 m} -  m c^2 ]\hat \phi - m c^2 \hat \phi^2 +h.c. \big \}~,
\end{split}
\end{equation}
where $\rho  $ gives the density of condensed atoms, $m$ the atom mass, and $c$ the (time dependent) speed of sound. The later is related to the time-dependent healing length by $\xi(t) c(t)= 1/2 m $.

The action for the environment degrees of freedom is taken of the form~\cite{Parentani:2007uq,Adamek:2008mp}
\begin{equation}
\begin{split}
S_\Psi &= \frac12 \int dt dx \int_{-\infty}^\infty d\zeta \left \{ \abs{\partial_t \hat \Psi_\zeta}^2 - \abs{\pi \zeta  \hat \Psi_\zeta}^2 \right \}~,
\label{SPsi}
\end{split}
\end{equation}
where the extra variable $\zeta$ has dimension of a frequency, and where $\hat \Psi_\zeta$ obeys $\hat \Psi_\zeta^\dagger =\hat \Psi_{-\zeta}$, as a hermitian field in momentum space. The variable $\zeta$  has been introduced in order to have infinitely more degrees of freedom in the $\hat \Psi$ field than in $\hat \phi$, a condition necessary to get dissipation~\cite{Unruh:1989dd}. (As shall be clear in the sequel, it is simpler to work with a continuous set, rather than an infinite discrete one as in Ref.~\cite{Caldeira:1981rx}.) Notice that $S_\psi$ contains no spatial gradients, and that it does not depend on $\rho$. Hence, the kinematics of the environment degrees of freedom is independent of both $k$ and $\rho$. This is a simplifying approximation. In fact, our philosophy is to choose the simplest model that possesses some key properties. These are: unitarity of the whole system, standard action for the phonons, and set of possibilities to describe dissipative properties. Given \eq{SPsi}, the choices are made in the third action, that governing the coupling. This action is taken to be 
\begin{equation}
\begin{split}
S_{\it int}\!  &=\!  - \! \!  \int \! \!  dt dx  \frac{g}{\sqrt{\xi}}  \left \{  (\rho \xi)^\alpha \,  (\xi \partial_x)^n (\hat \phi \! +\!  \hat \phi^\dagger) \, \partial_t (\int \! \! d\zeta \hat \Psi_\zeta) \right \}~,
\label{Sint}
\end{split}
\end{equation}
where $g$ is dimensionless. The powers $\alpha$ and $n$ imply that the strength of the coupling can vary with the density of condensed atoms and the wave number $k$. This is to account for the fact that different media behave differently in these respects. 

\subsection{Field equations and effective dispersion relation}

Because the condensate is homogeneous, it is appropriate to work with the Fourier components at fixed wave vector $\bk = - i\partial_x$, where $\bk$ is real. Then the total action splits into sectors that do not interact:  $S_T = \int d\bk S_\bk$, with $S_\bk = S_{-\bk}^\dagger$. In the rest frame of the condensate, at fixed $\bk$, the field equations are, with $k = \abs{\bk}$ 
\begin{subequations}
\label{eq:eomphipsi}
\begin{align}
\label{eq:eomphi}
[i\partial_t -\frac{k^2}{2m} -m c^2 ]\hat \phi_\bk &= m c^2 \hat \phi_{-\bk}^\dagger +  \frac{\gamma_\bk }{\sqrt{\rho}} \partial_t \int \! \! d\zeta \hat \Psi_{\zeta,\bk}~, \\
\label{eq:eompsi}
[\partial_t^2 + (\pi \zeta )^2 ] \hat \Psi_{\zeta,\bk} &=  \partial_t\left \{  \gamma_\bk^* \sqrt{\rho}   \, (\hat \phi_\bk +  \hat \phi^\dagger_{-\bk})\right \} \doteq \hat {j}_{\Psi,\bk}~.
\end{align}
\end{subequations}
where 
\begin{equation}
\begin{split}
 \gamma_\bk &= g (\rho \xi)^{\alpha - 1/2}\, (i \xi \bk)^n ~,  
\label{gamma}
\end{split}
\end{equation}
is the effective dimensionless coupling for the wave number $k$. The solution of \eq{eq:eompsi} is
\begin{equation}
\label{eq:psidecom}
\begin{split}
\hat \Psi_{\zeta,\bk} = \hat \Psi_{\zeta,\bk}^0 + \int dt' R^0_\zeta(t,t') \, \hat{ j}_{\Psi,\bk}(t')~,
\end{split}
\end{equation}
where $\hat \Psi_{\zeta,\bk}^0$ is an homogeneous solution we shall describe below, and where $R^0_\zeta$ is the retarded Green function. It is independent of $\bk $ because the action $S_\Psi$ contains no spatial gradient. When summing over $\zeta$, it obeys~\cite{Unruh:1989dd,Parentani:2007uq} 
\begin{equation}
\label{eq:localgretofpsi}
\begin{split}
\partial_t \left (\int d\zeta  R^0_\zeta(t,t') \right )  = \delta(t-t')~.
\end{split}
\end{equation}
This guarantees that the kernel encoding dissipation is local. Indeed, inserting Eq.~\eqref{eq:psidecom} into Eq.~\eqref{eq:eomphi}, and using \eq{eq:localgretofpsi}, Eq.~\eqref{eq:eomphi} gives 
\begin{equation}
\label{eqofmot}
\begin{split}
 [i \partial_t  - \frac{k^2}{2m} -m c^2 ]\hat \phi_\bk &=  m c^2 \hat \phi_{-\bk}^\dagger   +  \gamma_\bk  \partial_t \hat \Psi_\bk^0 /\sqrt{ \rho}  \\
 &\quad +  \gamma_\bk \partial_t\left \{ \gamma_\bk^*  (\hat \phi_\bk + \hat \phi_{-\bk}^\dagger)\right \} ~,
\end{split}
\end{equation}
where $\hat \Psi_\bk^0 = \int d\zeta \hat \Psi_{\zeta,\bk}^0$. As announced, the last term in the r.h.s.\ is local in time. Basically all other choices of $S_\Psi$ and $S_{\it int}$ would give a non-local kernel. In these cases, \eq{eqofmot} would be an integralo-differential equation. These more complicated models do not seem appropriate to efficiently calculate the consequences of dissipation on phonon correlation functions. 

To get the effective dispersion relation, we consider \eq{eqofmot} when all background quantities are constant and when $\Psi^0_\bk = 0$. We get, as in Eq.~\eqref{dr} 
\begin{equation}
\label{eq:disprel}
\begin{split}
& (\Omega_k + i \Gamma_k )^2 = \omega_k^2 ~,
\end{split}
\end{equation}
where 
\begin{equation}
\label{omGamma}
\begin{split}
  \omega_k^2 &= c^2 k^2 ( 1+  \xi^2 k^2) - \Gamma_k^2~, \\
  \Gamma_k    &= \abs{\gamma_\bk}^2\,  k^2 c \xi  ~. 
\end{split}
\end{equation}

Using \eq{gamma}, we verify that the second line delivers \eq{Gammapheno}. Hence, by choosing $g$, $\alpha$ and $n$ our model shall be able to reproduce many ab initio computed decay rates. As expected, we also verify that for $\abs{\gamma_\bk}^2 \to 0$, one recovers the standard Bogoliubov dispersion for all $k$. Our models thus provide dissipative extensions of some dispersive model. 

In what follows, the quantities $g,c,\xi$ depend on time, while preserving the homogeneity. Hence the non-trivial dynamics will occur within two mode sectors $\{\bk, - \bk\}$. 

\subsection{Time dependent settings}

For homogeneous time dependent systems, it is appropriate to introduce the auxiliary field 
\begin{equation}
\label{defchi}
\hat \chi_\bk \doteq - \frac{\hat \phi_\bk + \hat \phi^\dagger_{-\bk}}{\sqrt{2}} \sqrt{\frac{\rho}{ c \xi k^2}}~.
\end{equation}
Using Eq.~\eqref{eqofmot}, its time derivative is given by 
\begin{equation}
\partial_t \hat \chi_\bk = i \sqrt{\rho\xi  c  k^2 } \frac{\hat \phi_\bk - \hat \phi^\dagger_{-\bk}}{\sqrt{2}}~.
\end{equation}
The $\hat \chi$ field is both hermitian ($\hat \chi_\bk^\dagger  = \hat \chi_{-\bk}$) and canonical: it verifies the equal time commutators (ETC) $[\hat \chi_\bk,\partial_t\hat \chi_{\bk}^\dagger ] = i , [\hat \chi_\bk,\hat \chi_{\bk}^\dagger] = 0$, which are the usual ones for a relativistic scalar field in $k$ space. Moreover, $\hat \chi_\bk$ is simply related to the relative density fluctuation, and to the phase fluctuation, see Appendix~\ref{app:deltarhog1g2}. Finally, $S_\bk$, the action of the $\bk$ sector, reads 
\begin{equation}
\label{eq:Sofchi}
\begin{split}
S_\bk= \frac{1}{2} \int dt &\Big\{\vert \partial_t \hat \chi_\bk \vert^2  - (ck)^2 [1 + (k\xi)^2] \vert \hat \chi_\bk \vert^2  \\
& + \int d\zeta {\vert \partial_t \hat \Psi_{\zeta,\bk}\vert^2 } - (\pi \zeta)^2 \vert {\hat \Psi_{\zeta,\bk} \vert^2 }\\
& + 2  \hat \chi^\dagger_{\bk} \sqrt{2 \Gamma_k}  \partial_t \int d\zeta\hat \Psi_{\zeta,\bk} \Big\}~ ,
\end{split}
\end{equation}
where $c$, $\xi$,  and $\Gamma$ are arbitrary (positive) time dependent functions, and where a phase $( -i sgn(\bk))^n$ has been absorbed in $ \hat \Psi_{\zeta,\bk}$. In an atomic condensate, $c$ and $\xi$ are related by $c \xi = 1/2m = cst.$ We can also make the analogy with field propagation in a homogeneous cosmological background~\cite{Nation:2011uv,Carusotto:2009re}. Indeed, $c(t)$ acts as the inverse of the scale factor $a(t)$ (and not as a varying speed of light). Hence, a decreasing speed of sound corresponds to an expanding universe.

From the above action, or from Eq.~\eqref{eqofmot}, we get the equation for $\chi_\bk$:
\begin{equation}
\label{eomchi}
\left [ (\partial_t + \Gamma_k)^2 + \omega_k^2\right ]  \hat \chi_\bk = \sqrt{2 \Gamma_k }\partial_t \hat \Psi_\bk^0 ~.
\end{equation}
The general solution can be written as
\begin{equation}
\begin{split}
\hat \chi_\bk(t) &= \hat \chi_\bk^{dec}(t;t_0) + \hat  \chi_\bk^{dr}(t;t_0)~, 
\label{decomp}
\end{split}
\end{equation}
where the driven part $\hat \chi_\bk^{dr}(t;t_0)$, and its temporal derivative, vanish at $t=t_0$. The decaying part $\hat \chi_\bk^{dec}(t;t_0)$ is thus the solution of the homogeneous equation which obeys the ETC at that time. Hence it possesses the following decomposition
\begin{equation}
\label{chidec}
\begin{split}
\hat \chi_\bk^{dec}(t;t_0) &= \ep{- \int_{t_0}^t\Gamma_k dt' } \left ( \hat a_\bk \varphi_k(t) +  \hat a^\dagger_{-\bk} \varphi_{k}^*(t) \right )~, 
\end{split}
\end{equation}
where the destruction and creation operators $\hat a_\bk,\, \hat a^\dagger_{-\bk}$ obey the standard canonical commutators $[\hat a_{\bk},\hat a^\dagger_\bk]=1$, and where $\varphi_k$ is a solution of
 \begin{equation}
 \label{eq:homoeqvarphi}
\begin{split}
(\partial_t ^2 + \omega_k^2 ) \varphi_k = 0~, 
\end{split}
\end{equation}
of unit Wronskian $i(\varphi_k^* \partial_t \varphi_k -\varphi_k \partial_t \varphi_k^* ) =1$. The usefulness of this decomposition is two fold. On the one hand, $t_0$ can be conceived as the initial time when the state is fixed. The operators $\hat a^\dagger_{\bk},\hat a_\bk $ can then be used to specify the particle content of this state. On the other hand, \eq{decomp} and \eq{chidec} furnish an \enquote{instantaneous} particle representation around any time $t_0$. Indeed, in the limit $\Gamma/\omega \ll 1$ and $\Gamma (t-t_0)\ll 1$, the contribution of $\hat  \chi_\bk^{dr}(t;t_0)$ can be neglected, and $\hat \chi_\bk (t)\sim \hat \chi_\bk^{dec}(t;t_0)$ behaves as a standard canonical field since the prefactor of Eq.~\eqref{chidec} is approximatively equal to $1$. We shall return to this in Sec.~\ref{sec:nbofpartdissip}.

The driven part of Eq.~\eqref{decomp} is given by
\begin{equation}
\label{chidriven}
\begin{split}
\hat \chi_\bk^{dr}(t,t_0) = \int_{t_0}^\infty dt' G_{ret} (t,t';k) \sqrt{2 \Gamma_k(t') } \partial_{t'} \hat \Psi_\bk^0(t')~,
\end{split}
\end{equation}
where $G_{ret}$ is the retarded Green function of Eq.~\eqref{eomchi}. Using the unit Wronskian solution of \eq{eq:homoeqvarphi}, it can be expressed as
\begin{equation}
\label{Gret}
\begin{split}
G_{ret}(t,t';k) = \theta (t - t')\,  \ep{-\int_{t'}^{t} \Gamma_k dt} \times 2\, \Im(\varphi_k(t)\varphi_k^*(t')) ~.
\end{split}
\end{equation}

Since $\hat \chi$ is a canonical and linear field, the standard relation between the commutator and the retarded Green function holds, namely
\begin{equation}
[\hat \chi_\bk(t),\hat \chi_{-\bk}(t')] \theta(t-t') = i\,  G_{ret}(t,t';k)~.
\end{equation} 
In consequence, when the state of the system is Gaussian and homogeneous, the reduced state of $\chi$ is completely fixed by its anti-commutator. Because of Eq.~\eqref{decomp}, it contains 3 terms
\begin{equation}
\label{Gact0}
\begin{split}
G_{ac}(t,t';k ) &\doteq {\rm Tr}(\hat \rho_T \, \{\hat \chi_\bk(t), \hat \chi_{-\bk}(t')\}) \\
&= G_{ac}^{dec} +G_{ac}^{dr} +G_{ac}^{mix} ~.
\end{split}
\end{equation}
The first one decays and is governed by $\hat \chi^{dec}$
\begin{equation}
\label{Gacdec}
\begin{split}
G_{ac}^{dec}(t,t';k) = {\rm Tr}(\hat \rho_T \{\hat \chi_\bk^{dec}(t), \hat \chi_{-\bk}^{dec}(t')\})~.
\end{split}
\end{equation}
The second one is driven and governed by $\hat \Psi^0$
\begin{equation}
\label{eq:Gacasintegrals}
\begin{split}
G_{ac}^{dr}(t,t';k) = &\int_{t_0}^\infty  d\tau d\tau' \sqrt{2\Gamma_k(\tau)}\sqrt{2\Gamma_k(\tau')}G_{ret}(t,\tau)\\
& G_{ret}(t',\tau' )  \partial_\tau \partial_{\tau'} \rm{Tr}(\hat \rho_T \{\hat \Psi_\bk^0(\tau), \hat \Psi_{-\bk}^0(\tau')\})~.
\end{split}
\end{equation}
The third one describes the correlations between $\chi$ and $\Psi$. It is non zero either when the initial state is not factorized as $\hat \rho = \hat \rho_\chi \otimes \hat \rho_\Psi$, or when the two fields have interacted. It is given by twice the symmetrized of 
\begin{equation}
\label{Gacmix}
\begin{split}
\widetilde G_{ac}^{mix}(t,t';k) \doteq \!  \int_{t_0}^\infty \! \! \!  d\tau & \sqrt{2\Gamma_k(\tau) }G_{ret}(t',\tau) \\
 & \partial_\tau {\rm Tr}( \hat \rho_T \{ \hat \chi_\bk^{dec}(t),\hat \Psi_{-\bk}^0(\tau)\})~.
\end{split}
\end{equation}
When the state is prepared at an early time $ \Gamma (t -t_0)\gg 1$, only the driven term significantly contributes to Eq.~\eqref{Gact0}, which means that the system would have thermalized with the bath. 

At fixed $\bk$ and $\zeta$,  $\hat \Psi_{\zeta,\bk}^0$, the homogeneous solution of Eq.~\eqref{eq:eomphipsi} is a complex harmonic oscillator of pulsation $ \omega_\zeta = \pi  \abs{\zeta}$. Hence it can be expressed as
\begin{equation}
\label{eq:psi0fock}
\begin{split}
\hat \Psi_{\zeta, \bk}^0 (t) &=  \frac{\ep{- i \omega_\zeta t} \hat a_{\zeta, \bk} +\ep{ i \omega_\zeta t} \hat a_{-\zeta, -\bk}^\dagger }{\sqrt{2  \omega_\zeta}}~, 
\end{split}
\end{equation}
where $\hat a_{\zeta, \bk}$ and $\hat a_{\zeta,\bk}^\dagger $ are standard destruction and creation operators. In the following Sections we shall work with thermal baths at temperature $T$. In such states, the noise kernel entering Eq.~\eqref{eq:Gacasintegrals} is, with Boltzmann constant $ k_B = 1$, 
\begin{equation}
\begin{split}
\label{eq:Trpsipsi}
{\rm Tr}(\hat \rho_T & \{\hat \Psi_\bk ^0(\tau),\hat  \Psi_{-\bk}^0(\tau')\})  \\
&= \int_{0}^\infty \!  \frac{d\omega_\zeta }{ \pi } \frac{\coth\left (\frac{\omega_\zeta}{2 T}
\right )}{2 \omega_\zeta} \cos[\omega_\zeta (\tau - \tau')] ~.
\end{split}
\end{equation}
We notice that it does not depend on $k$. 

\subsection{Sudden changes}
\label{sec:gradino}

We study the time evolution of the phonon state when making two assumptions. First, we consider states which are prepared a long time before the experiment, so that Eq.~\eqref{Gact0} is given by Eq.~\eqref{eq:Gacasintegrals}, with $t_0 = -\infty$. Second, we suppose a sudden change of the condensed atoms occurs at time $t=0$. Hence, the speed of sound $c$ and the effective coupling $\gamma_\bk$ entering Eq.~\eqref{eomchi} will change on a similar time scale. Since approximating the change of the sound speed by a step function only modifies the response for very high $k$, as can be seen in Ref.~\cite{Carusotto:2009re}, for simplicity, we shall work with an instantaneous change for $c$. For $\gamma_\bk$ instead, we shall use a continuous profile because an instantaneous change would lead to divergences, as we shall see below. Hence we shall work with
\begin{equation}
\label{eq:grad}
\begin{split}
 c(t) &=c_{\rm f} + (c_{\rm in} -c_{\rm f})\, \theta(-t) ~,\\
 \gamma_\bk (t) &=  \gamma_{\rm f} + (\gamma_{\rm in} - \gamma_{\rm f})\, h(\kappa t)~, 
\end{split}
\end{equation}
where $h(\kappa t)$ is a smoothing out function which goes from $1$ to $0$ around $t= 0$ in a time lapse of the order of $1/\kappa$. 
Considering the asymptotic values of these profiles, we shall use the following notations
\begin{equation}
\label{eq:defGammaomegainandf}
\begin{split}
\Gamma_{\rm in / f} &\doteq \gamma_{\rm in / f}^2 \, (c \xi)\,  k^2 ~,\\
\omega_{\rm in / f}^2 &\doteq c_{\rm in / f}^2 k^2 + (c \xi)^2 k^4 - \Gamma_{\rm in / f}^2~, 
\end{split}
\end{equation}
see Eq.~\eqref{omGamma}. 

Before considering dissipation, we study the dispersive case, firstly, to analyze the reduction of the coherence due to stimulated processes, and secondly, to know the outcome in the dissipation free case, so as to be able to isolate the consequences of dissipation.

\section{The dispersive case}
\label{sec:nodiss}

\begin{figure}
\includegraphics[width=8cm]{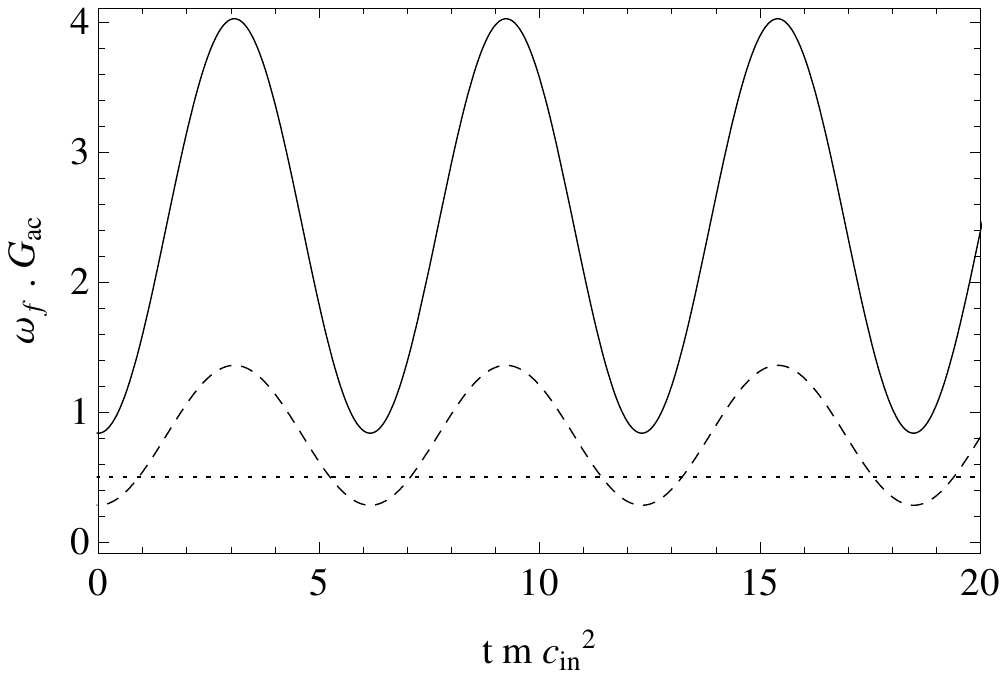}
\caption{We represent the product $\omega_{\rm f} \times G_{ac}(t,t'=t)$ where $G_{ac}$ is given in Eq.~\eqref{eq:gacofnc} as a function of the adimensionalized time $t m c_{\rm in}^2$, for $ k = m c_{\rm in} $, and for two values of the temperature, namely $ T_{\xi_{\rm in}} /2 $ (dashed) and $ 2 T_{\xi_{\rm in}}  $ (solid), see \eq{eq:defTxiin}. The value of the jump is $c_{\rm f}/ c_{\rm in} = 0.1$. As explained in the text, the dotted line gives the threshold value $1/2$ which distinguishes non-separable states. When increasing the temperature, the contribution of the stimulated amplification with respect to the spontaneous one is larger. As a result, the coherence is reduced, i.e. the minima of $\omega_{\rm f} \times G_{ac} $ are increased.} 
\label{fig:Gacoft}
\end{figure}

\begin{figure*}[htb] 
\begin{minipage}{0.47\linewidth}
\includegraphics[width=1 \linewidth]{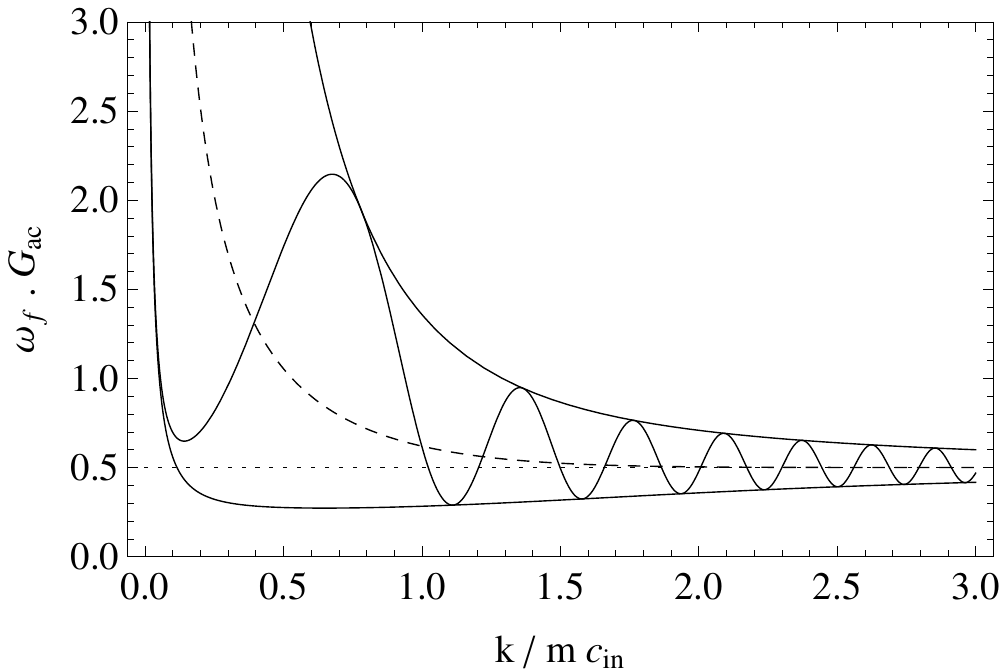}
\end{minipage}
\hspace{0.03\linewidth}
\begin{minipage}{0.47\linewidth}
\includegraphics[width=1 \linewidth]{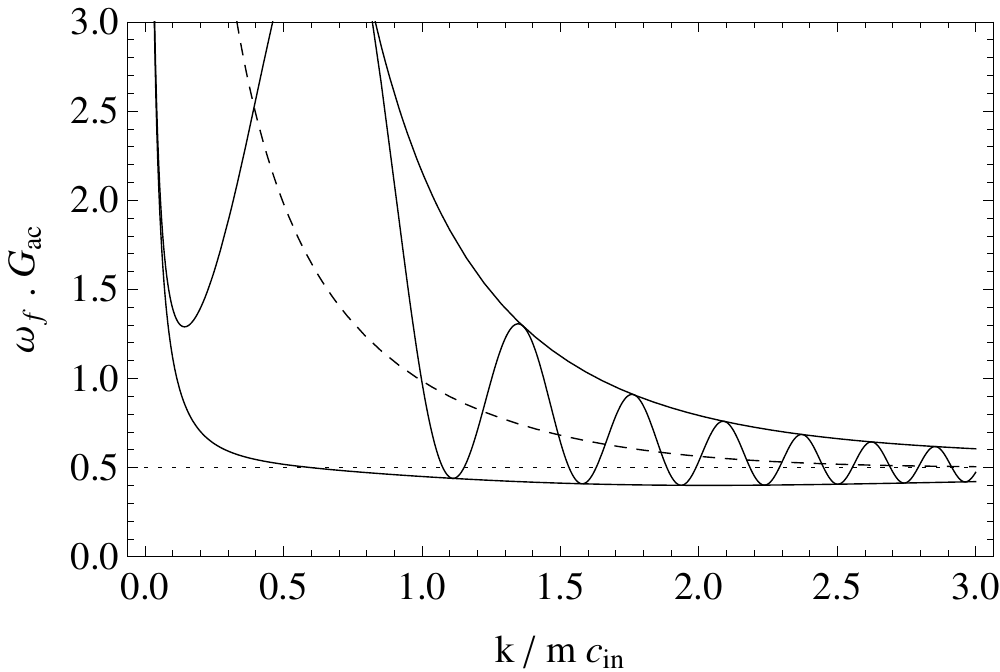}
\end{minipage}
\caption{The anticommutator of Eq.~\eqref{eq:gacofnc} multiplied by $\omega_{\rm f}$ as a function of $k /m c_{\rm in}$, when evaluated at equal time $t=t' = 5 / m c_{\rm in}^2$, for $c_{\rm f} = c_{\rm in}/10$, on the left panel for $T = T_{\xi_{\rm in}}/2$, and on the right for $T=T_{\xi_{\rm in}}$ (solid oscillating curves). The dashed line in the middle gives the value of $\omega_{\rm f} G_{ac} = n^{in} + 1/2$ before the sudden change. The envelopes of the minima and maxima are also indicated in solid lines. One clearly sees that the domain of $k$ where the lower line is below the threshold value $1/2$ is reduced when increasing the temperature. Notice that all curves asymptote to $1/2$ because in the limit $k\to \infty$, one has $n_k=c_k=0$. }
\label{fig:Gacofk}
\end{figure*} 

\begin{figure*}
\begin{minipage}{0.43\linewidth}
\includegraphics[width=1\linewidth]{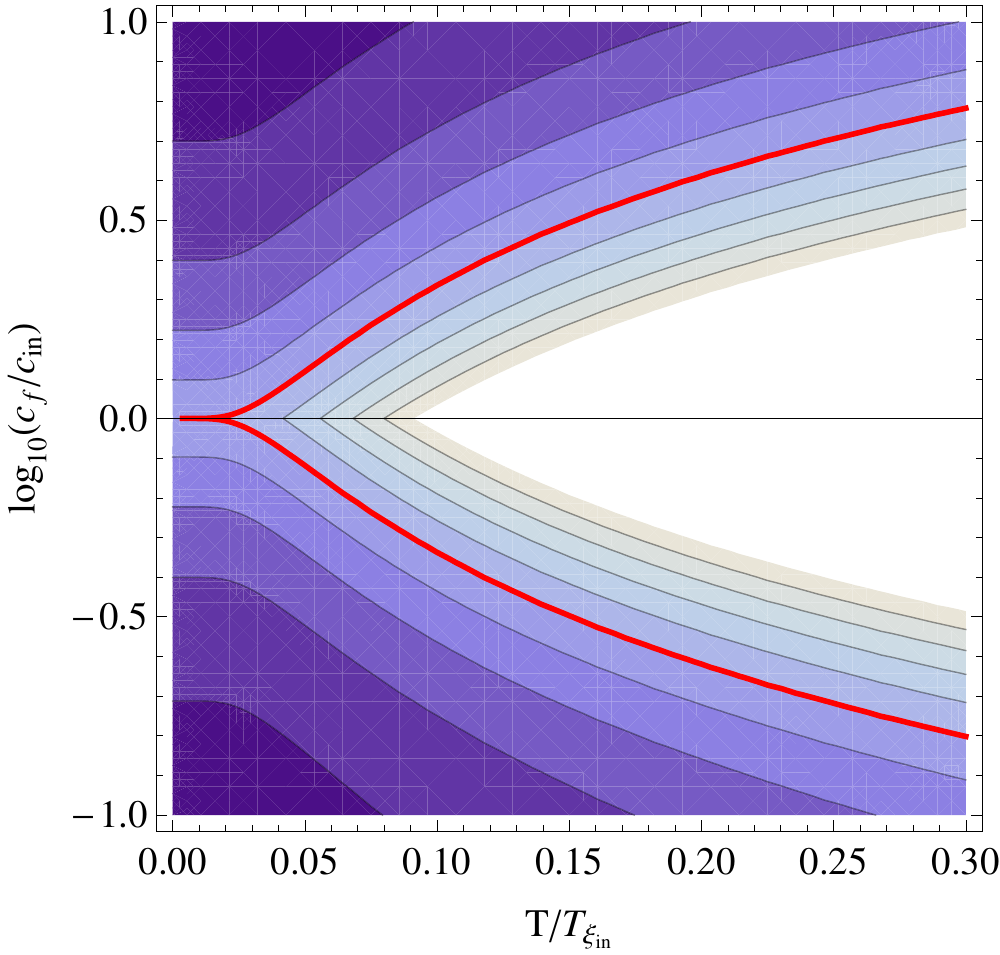}
\end{minipage}
\hspace{0.01\linewidth}
\begin{minipage}{0.09\linewidth}
\includegraphics[width = 1 \linewidth]{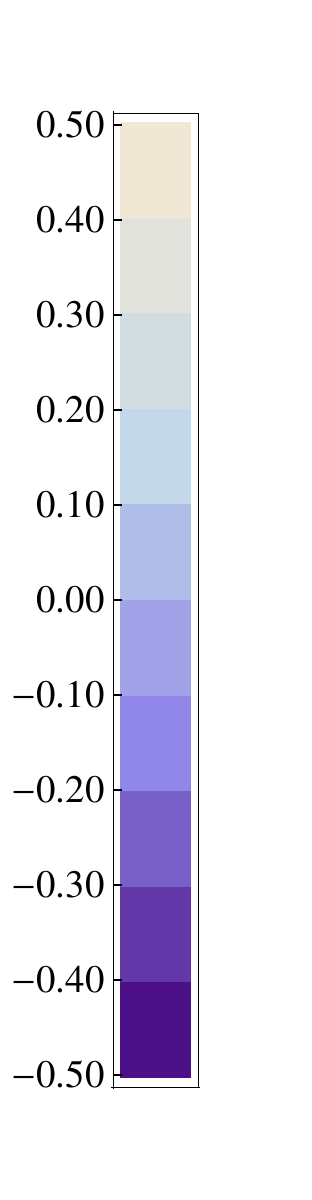}
\vspace{0.5cm}\ 
\end{minipage}
\begin{minipage}{0.43\linewidth}
\includegraphics[width=1\linewidth]{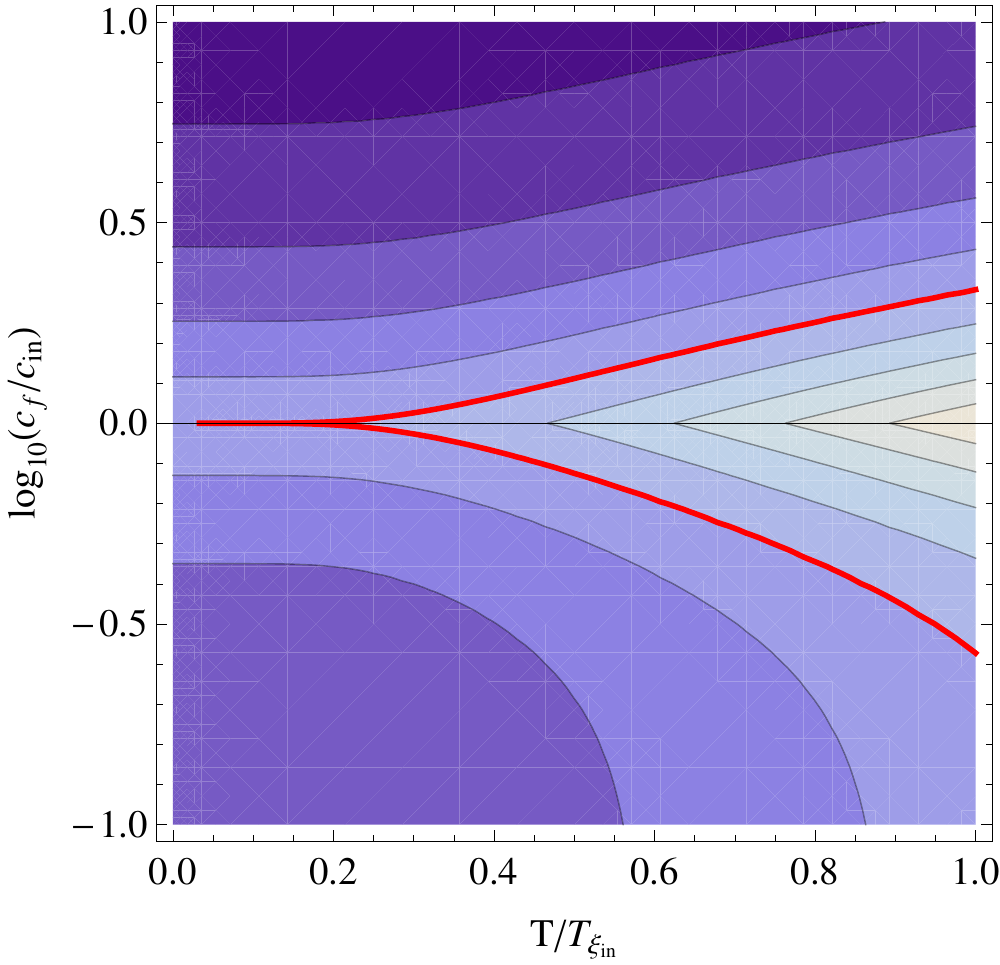} 
\end{minipage}
\caption{Contour plot of $\Delta^{out}$ induced by a sudden variation of the sound speed, as a function of temperature $T/T_{\xi_{\rm in}}$ and the logarithm of the ratio $c_{\rm f}/c_{\rm in}$. In the left panel, $k  = m c_{\rm in}/10$ is in the hydrodynamical regime, and in the right panel, $k  = m c_{\rm in}$. The threshold value $\Delta^{out} = 0$ is indicated by a thick line. One clearly sees the competition between the hight of the jump governed by $c_{\rm f}/c_{\rm in}$ which increases the coherence, i.e. reduces the value of $\Delta$, and the initial occupation number which increases $\Delta$. One also sees that the state of a higher momentum mode stays non separable for higher temperatures. }
\label{deltadispgradino1}
\end{figure*} 

In the absence of dissipation, phonon excitations can be analyzed before and after the jump using a standard particle interpretation. Hence, the consequences of the jump are all encoded in the Bogoliubov coefficients $\alpha,\beta$ entering
\begin{equation}
\begin{split}
\varphi_{in} &= \alpha \varphi_{out} + \beta \varphi_{out}^* ~,
\label{inout}
\end{split}
\end{equation}
which relates the $in$ mode to the $out$ mode. These modes have a positive unit Wronskian and are equal to $\ep{-i \omega t} \big/ \sqrt{2 \omega }$ for $t<0$ or $t>0$ respectively. Using Eq.~\eqref{eq:homoeqvarphi}, one verifies that the modes are $\mathcal{C}^1$ across the jump. From the junction conditions, one finds the Bogoliubov coefficients~\cite{Carusotto:2009re} 
\begin{equation}
\label{bogodisp}
\alpha = \frac{\omega_{\rm f}+\omega_{\rm in}}{2\sqrt{\omega_{\rm f} \omega_{\rm in}}}~, \quad \beta = \frac{\omega_{\rm f}-\omega_{\rm in}}{2\sqrt{\omega_{\rm f} \omega_{\rm in}}}~.
\end{equation}

To prepare the comparison with the dissipative case, the initial phonon state is taken to be a thermal bath at temperature $T$. This means that the initial mean occupation number is $n^{in} = 1/({\ep{\omega_{\rm in} / T} -1 })$ and that $c^{in}$, the initial correlation term between $\bk$ and $-\bk$, vanishes. After the jump, the mean occupation number and the correlation term are 
\begin{subequations}
\label{outnc} 
\begin{align}
n^{out}&= n^{out}_{\rm spont.}+ n^{out}_{\rm stim.} ~, \nonumber \\
       &= \abs{\beta}^2  + \frac{\abs{\alpha}^2 + \abs{\beta}^2 }{\ep{\omega_{\rm in} /T} -1 } ~,\\
c^{out}&= c^{out}_{\rm spont.}+ c^{out}_{\rm stim.} ~, \nonumber \\
       &= \alpha \beta +  \frac{2 \alpha \beta }{\ep{\omega_{\rm in} /T} -1 }~.
\end{align}
\end{subequations}
These two quantities completely fix the late time behavior of the anticommutator $ G_{ac}(t,t') = (n^{in}+1/2){\rm Re }\{\varphi_{in}(t) \varphi_{in}^*(t')\}$. In fact, for $t, t' > 0$, one has 
\begin{equation}
\label{eq:gacofnc}
\begin{split}
 G_{ac}(t,t') =&  (n^{out}+1/2) \frac{\cos \left ( \omega_{\rm f} (t-t')\right  ) }{\omega_{\rm f}}\\
               &+ Re \left (c^{out} \frac{ \ep{-i \omega_{\rm f} (t+t')}}{\omega_{\rm f}}\right )~.
\end{split}
\end{equation}
Using this expression, it is clear that the contribution of the stimulated amplification (weighted by $n^{in} = 1/({\ep{\omega_{\rm in} /T} -1 })$) and that of spontaneous processes are not easy to distinguish. As recalled in the Introduction, to be able to do so, it is useful to introduce the parameter $\Delta$ of \eq{deltak}. One can show that it obeys $-1/2 < n-\sqrt{n(n+1)} \leq\Delta \leq n$, where the minimal and maximum values characterize respectively the pure (squeezed) state, and the incoherent thermal state~\cite{Campo:2005sy}. In addition, whenever $\Delta < 0$, the two-mode state $\{ \bk, -\bk\} $ is non-separable in the sense of Werner~\cite{Werner:1989}.\footnote{ $\Delta$ is linked to the logarithmic negativity $E_\mathcal{N}$ introduced in Ref.~\cite{Vidal:2002zz} and used in Ref.~\cite{Horstmann:2009yh,Horstmann:2010xd}. Indeed, in our case, one finds $E_\mathcal{N} = \max[-\log_2(1+ 2\Delta),0]$ which is positive only if $\Delta <0$. }
This means that the correlations are so strong that they cannot be described by a classical ensemble. For a recent application using dissipative fields in a cosmological context, we refer to~\cite{Adamek:2013vw}. In the present case, the value of $\Delta$ associated to Eqs.~\eqref{outnc} is
\begin{equation}
\begin{split}
\Delta^{out}&= \Delta^{out}_{\rm spont.}+ \Delta^{out}_{\rm stim.} ~, \\
            &= -(\abs{\alpha}- \abs{\beta}) \abs{\beta}  + \frac{(\abs{\alpha} - \abs{\beta})^2 }{\ep{\omega_{\rm in} /T} -1 }~.
\label{deltaout}
\end{split}
\end{equation}
At fixed $\abs{\beta/\alpha}$, the threshold value of non-separability $\Delta=0$ defines a critical temperature $T_C$. It is given by
\begin{equation}
\label{eq:Tcdef}
\begin{split}
\abs{\beta/\alpha} = \ep{- \omega_{\rm in}/ T_C} ~. 
\end{split}
\end{equation}
In \eq{deltaout}, one clearly sees the competition between the squeezing of the state due to the sudden jump governed by $\beta$ which reduces the value of $\Delta$, and the initial occupation number which increases its value. It remains to extract these informations from \eq{eq:gacofnc}. 

To this end, we plot in Fig.~\ref{fig:Gacoft} the product $\omega_{\rm f} \, G_{ac}(t,t'=t)$ of Eq.~\eqref{eq:gacofnc} as a function of time, for two different values of $T$, respectively half and twice the temperature 
\begin{equation}
\label{eq:defTxiin}
\begin{split}
 T_{\xi_{\rm in}} \doteq  m c_{\rm in}^2 = \frac{1}{4 m \xi_{\rm in}^2} ~, 
\end{split}
\end{equation} 
fixed by the initial value of the healing length. We obtain two perfect sinusoidal curves since the mode $\chi_k$ freely propagates after the sudden change. Interestingly, we can read the values of $n$ and $\abs c$ from the envelope of the curve. Indeed, the maxima reach $n +1/2 + \abs{c} $ and the minima $ n +1/2 - \abs{c} = \Delta +1/2$. Therefore, if in an experiment, the minimal value of $\omega_{\rm f}  G_{ac}$ is measured with enough precision to be less than $1/2$, one can assert that the state is non-separable (in the absence of dissipation). This identification can also be obtained using the so-called $g_1$ correlation function, or the $g_2$, see appendix~\ref{app:deltarhog1g2}. One could also work at fixed $t$ and vary $t'$. In this case, one would get a periodic behavior of frequency $\omega_{\rm f} /2\pi$. However, the curve is now centered on $0$, and the maxima vary from $n+1/2 -\abs c$ to $n+1/2 +\abs c$ according to the value of the fixed time $t$. Hence, the non separability of the state reveals itself in the fact that there exists some values of $t$ such that $\omega_{\rm f}  G_{ac} (t, t')$ remains smaller than $1/2$ for all $t'$. 

To complete the analysis, we study how these results depend on the wave number $k$. We work with the standard Bogoliubov dispersion relation, see Eq.~\eqref{eq:disprel} with $\Gamma = 0$, $ c \xi = 1/2 m$ constant, and with $c_{\rm f}/c_{\rm in} = 1/10$. In Fig.~\ref{fig:Gacofk} we plot $\omega_{\rm f} \times G_{ac}(t,t'=t)$ as a function of $k$ for two temperatures, namely $T_{\xi_{\rm in}}/2$ (left panel) and $T_{\xi_{\rm in}}$ (right panel). We first see that the modes with lower $k$ are more amplified than those with higher $k$. As expected from Eq.~\eqref{eq:Tcdef}, when looking at the lower envelope, we also see that the coherence level is higher (the minima of $G_{ac}$ lower) when working with a smaller temperature, and/or with higher $k$, i.e., with rarer events governed by a smaller initial occupation number $n_{in}$. To quantify this effect, and possibly also to guide future experiments, we characterize the domain where the resulting state is non-separable, i.e., where $\Delta^{out} < 0$. To this end, in Fig.~\ref{deltadispgradino1}, we plot $\Delta^{out}$ as a function of $T/T_{\xi_{\rm in}}$ and the ratio $c_{\rm f}/c_{\rm in}$. We consider two values of $k$, namely one in the hydrodynamical regime, and one of the order of the inverse healing length. This clearly confirms that at higher momenta, states are more likely to be non separable. Moreover, one sees that for a wave number smaller than the healing length, at a temperature $\sim T_{\xi_{\rm in}}$ of \eq{eq:defTxiin}, in order to obtain a non separable state, $c_{\rm f} / c_{\rm in}$ should be either larger than $3$, or smaller than $1/3$.

To illustrate these aspects with a concrete example, we consider the experiment of Ref.~\cite{PhysRevLett.109.220401}. The relevant values are $T  =  6.05 T_{\xi_{\rm in}}$ and $k \sim 2.15 m c_{\rm in}$, so that the initial number of particles is of the order of $3$. On the other hand, one has $c_{\rm f} / c_{\rm in}  \sim 2^{1/4}$. (To get these numbers, we used $T=200nK$, $\omega/2\pi = 2kHz$, $m= 7. \,  10^{-27} kg$ and $c_{\rm in}=8mm/s$.) The corresponding value of the coherence level is $ \Delta \sim 1.4 $. Hence the state is separable. In order to reach $\Delta =0$, one should either increase the ratio $c_{\rm f} / c_{\rm in}  \sim 6 $, or work with a lower temperature, of the order of $0.6 T_{\xi_{\rm in}}$. 

\section{The dissipative case}
\label{sec:dissipative}

In presence of dissipation, the mode interpretation involving the Bogoliubov coefficients of \eq{inout} is no longer valid. In fact, the state of $\chi$ is now characterized by $G_{ac}^{dr}$ of Eq.~\eqref{eq:Gacasintegrals}, which is governed by the retarded Green function and the noise kernel. The separability of the state should thus be deduced from its properties.

When the environment state is a thermal state, using Eq.~\eqref{eq:Trpsipsi}, Eq.~\eqref{eq:Gacasintegrals} can be expressed as 
\begin{equation}
\label{Gacdr}
\begin{split}
G_{ac}^{dr}(t,t')& =  \int \frac{d\omega_\zeta}{2\pi}  \omega_\zeta  \coth\left (\frac{\omega_\zeta}{2T}\right )  \widetilde G_{\rm r}(t,\omega_\zeta)  \widetilde G_{\rm r}(t',-\omega_\zeta) ~,  
\end{split}
\end{equation}
where we introduced the Fourier transform 
\begin{equation}
\label{buildingblock}
\begin{split}
\widetilde G_{\rm r}(t,\omega_\zeta)&\doteq \int_{- \infty }^\infty d\tau \, \ep{i \omega_\zeta \tau} \sqrt{\Gamma(\tau)}\, G_{ret}(t,\tau)~,\\
\end{split}
\end{equation}
of the retarded Green function of \eq{Gret}. In the following, we compute \eq{Gacdr}, which is more easy to handle than Eq.~\eqref{eq:Gacasintegrals}, in two different cases. In the first one, there is no dissipation after the sudden change, i.e., $\gamma_{\rm f}=0$ in Eq.~\eqref{eq:grad}. In the second one, $\Gamma$ is constant.

\subsection{Turning off dissipation after the jump}

When $\Gamma_{\rm f}=0$ for $\kappa  t \gg 1$, we have the possibility of using the standard particle interpretation to read the asymptotic state. In fact, inserting Eq.~\eqref{Gret} in Eq.~\eqref{buildingblock}, using Eq.~\eqref{eq:grad} and $\kappa  t \gg 1$, one gets 
\begin{equation}
\label{eq:TFGretGammaf0}
\begin{split}
\widetilde G_{\rm r}(t,\omega_\zeta) = \frac{ \sqrt{\Gamma_{\rm in}}}{2 \omega_{\rm f}} \left [ \ep{i\omega_{\rm f} t} R( \omega_\zeta) - \ep{-i\omega_{\rm f} t}R^*( -\omega_\zeta) \right ]~,
\end{split} 
\end{equation}
where
\begin{equation}
\label{eq:defR}
\begin{split}
R(\omega_\zeta) \doteq \sqrt{2 \omega_{\rm f}} \int_{- \infty }^\infty d\tau h(\kappa \tau ) \ep{i \omega_{\zeta} \tau } \ep{ - \int_\tau^\infty \Gamma } \varphi^{out}~.
\end{split}
\end{equation}
The function $\varphi^{out}( \tau)$ is the standard out mode: it is the positive unit Wronskian mode of Eq.~\eqref{eq:homoeqvarphi} which is positive frequency at asymptotically late time. The time dependence of Eq.~\eqref{eq:TFGretGammaf0} guarantees that Eq.~\eqref{Gacdr} has exactly the form of Eq.~\eqref{eq:gacofnc}. The final occupation number $n^{out}$ and correlations $c^{out}$ are found to be
\begin{equation}
\label{eq:nandcdiverge}
\begin{split}
& n^{out} + \frac{1}{2} = \frac{\Gamma_{\rm in}}{\omega_{\rm f}} \int_{0}^\infty  \frac{ d\omega_\zeta}{\pi}  \omega_\zeta \coth (\frac{\omega_\zeta}{2T}) \\
& \hspace{2.8 cm } \times \left (\abs{R(\omega_\zeta)}^2+\abs{R(-\omega_\zeta)}^2 \right )  ~, \\ 
& c^{out\,*} = 2\frac{ \Gamma_{\rm in} }{\omega_{\rm f}} \int_{0}^\infty  \frac{ d\omega_\zeta}{\pi}   \omega_\zeta \coth (\frac{\omega_\zeta}{2 T}) R(\omega_\zeta) R(-\omega_\zeta)  ~.
\end{split}
\end{equation}
We notice that these expressions are similar to those of Eq.~\eqref{outnc}, and that $R^*(\pm \omega_\zeta)$ play the role of a density (in $\omega_\zeta$) of $\alpha$ and $\beta$ respectively. In fact, when taking the limit $\Gamma_{\rm in} \to 0$ in \eq{eq:nandcdiverge}, one recovers the dispersive expressions of Eq.~\eqref{outnc}. 

We can now explain why we introduced the function $h$ in \eq{eq:grad}. For $\kappa\to \infty$, $h(\kappa t)$ becomes the step function $\theta(-t)$. In this limit, \eq{eq:defR} gives 
\begin{equation}
R(\omega_\zeta) = \frac{ \omega_{\rm f} +( \omega_\zeta-i \Gamma_{\rm in}) }{\omega_{\rm in}^2 - ( \omega_\zeta- i \Gamma_{\rm in} )^2} + \mathcal{O } \left (\frac1\kappa\right )~,
\label{Rom}
\end{equation}
which indicates that $R$ behaves as $1/\omega_\zeta$ for $\omega_\zeta \to \infty$. Hence both $n^{out}$ and $c^{out}$ of Eq.~\eqref{eq:nandcdiverge} would logarithmically diverge. The divergences arise from the fact that the environment field $\hat \Psi$ contains arbitrary high frequencies $\omega_\zeta$. To regulate the divergences, several avenues can be envisaged. One could either introduce a $\zeta$ dependent coupling in \eq{Sint}, or cut off the high frequency $\omega_\zeta$ spectrum in \eq{SPsi}. However, these would spoil the locality of Eq.~\eqref{eq:localgretofpsi}. For this (mathematical) reason, we prefer to use $h(\kappa t)$ of \eq{eq:grad}. Moreover, taking an instantaneous change in $\Gamma(t)$ would remove the $\mathcal{C}^1$ character of $G_{ac}(t,t')$ found in the dispersive case, see the discussion after \eq{inout}. In addition it should be noticed that any physical system, such as an atomic Bose condensate described by Eq.~\eqref{grosspit}, would only respond after a finite amount of time. Hence the function $h$ of \eq{eq:grad} is physically meaningful, and $\kappa$ should be of the order of $c/\xi$ in an atomic condensate.

In Appendix~\ref{app:renorm2} we derive approximate expressions for $ R(\omega_\zeta)$, both for a general profile $h$, and when applied to the particular case
\begin{equation}
\label{eq:chosenf}
\begin{split}
h (z) = \left \{
\begin{array}{ll}
1 & \mbox{ if } z<0~,\\
1-z & \mbox{ if } 0<z<1~,\\
0 & \mbox{ if } 1<z~,
\end{array} \right .
\end{split}
\end{equation}
we shall use to obtain the following figures.

\begin{figure}
\begin{minipage}{0.75\linewidth} 
\includegraphics[width = 1\linewidth]{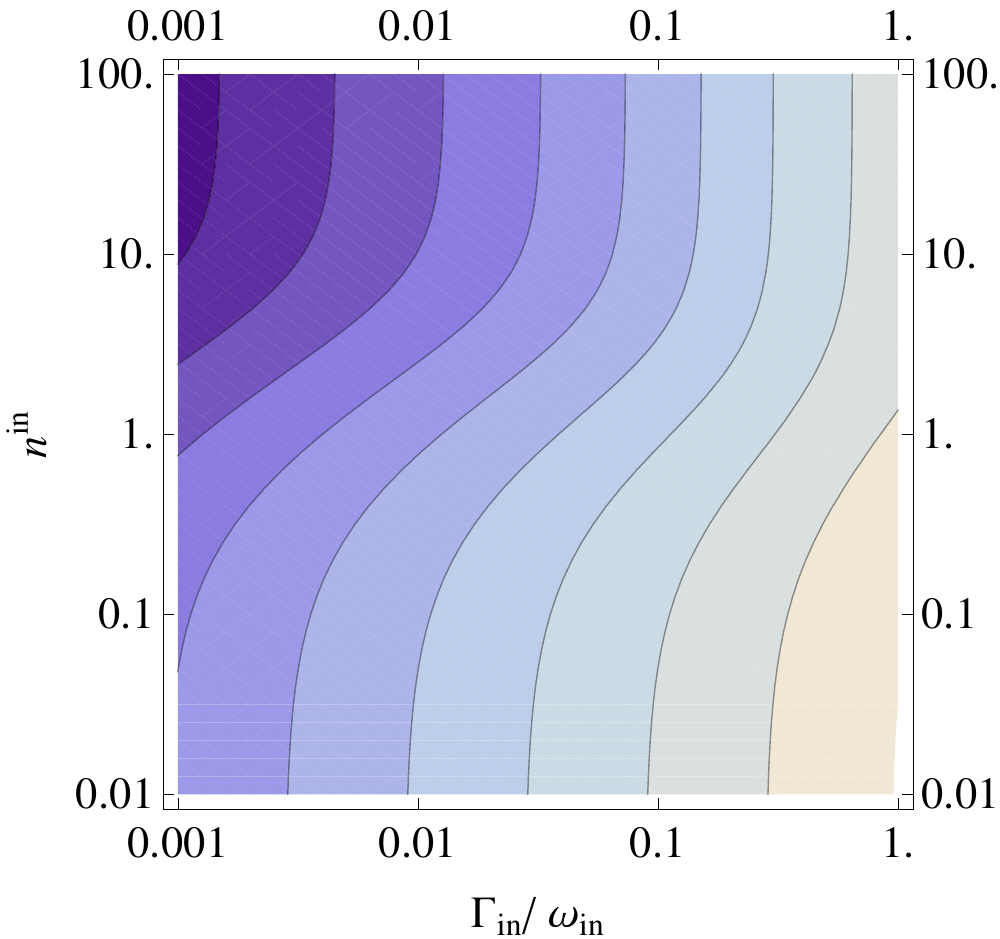} 
\end{minipage}
\begin{minipage}{0.2\linewidth}
\includegraphics[width = 1\linewidth]{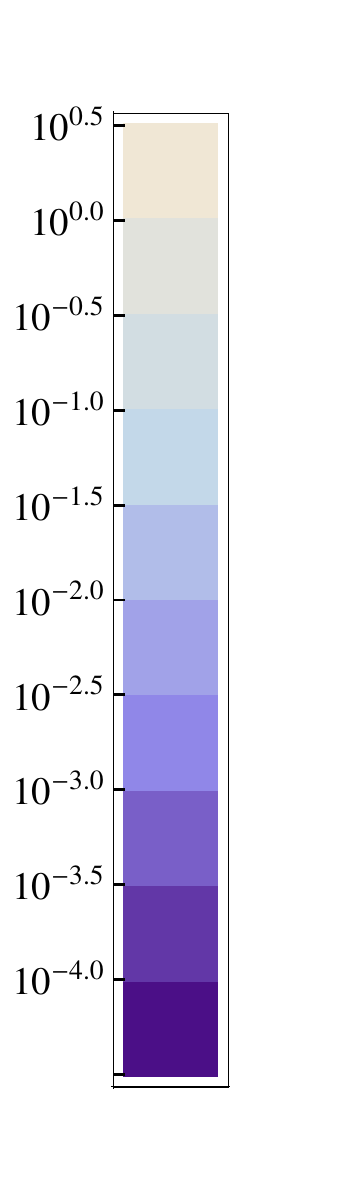}
\end{minipage}
\caption{The ratio $\Delta n_r$ of \eq{rapsp} for a jump $\omega_{\rm f} / \omega_{\rm in} = 0.1$ and $\kappa = 10 \omega_{\rm in}$ is represented in the plane $n^{in}, \, \Gamma_{\rm in}/ \omega_{\rm in}$. For low initial occupation numbers, one sees that $\Delta n_r $ is proportional to $\Gamma_{\rm in}/\omega_{\rm in} $, whereas it evolves from linear to quadratic for high numbers. As explained in the text, these observations are confirmed by analytical expressions.}
\label{fig:contourdeltanovern}
\end{figure}

\subsubsection{Spectral deviations due to dissipation}
\label{sec:nfirsteffet}

We study how $n^{out}$ of \eq{eq:nandcdiverge} depends on the dissipation rate $\Gamma_{\rm in}$. To this end, we study its difference with the dispersive occupation number $n^{out}_{disp}$ of Eq.~\eqref{outnc} evaluated with the same values for the temperature $T$, the initial and the final frequency, see Eq.~\eqref{eq:defGammaomegainandf}. In Fig.~\ref{fig:contourdeltanovern} we represent the relative change 
\begin{equation}
\begin{split}
\Delta n_r \doteq \frac{n^{out}(\Gamma_{\rm in}) - n^{out}_{disp}}{n^{out}_{disp}+1/2} ~, 
\label{rapsp}
\end{split}
\end{equation}
as a function of the initial occupation number $n^{ in}$ and the ratio $ \Gamma_{\rm in} / \omega_{\rm in}$. We work with a jump $\omega_{\rm f} / \omega_{\rm in} = 0.1$, and with $\kappa = 10 \omega_{\rm in}$. (We use this parameterization because, at fixed $k$, Eq.~\eqref{eomchi} only depends on $\omega(t)$ and $\Gamma(t)$. Hence the healing length and $k$ need not be specified.) For these values, we find two regimes. First, for low occupation number, we observe that the deviation $\Delta n_r$ linearly depends on $ \Gamma_{\rm in}$. An analytical treatment based on Eq.~\eqref{eq:Rforanyf} reveals that, for small $ \Gamma_{\rm in}/ \omega_{\rm in}$ and large $\kappa/ \omega_{\rm in}$, $\Delta n_r$ behaves as 
\begin{equation}
\begin{split}
\Delta n_r = \left (\frac{  \Gamma_{\rm in} }{\omega_{\rm in}}\right) \frac{1} {n^{in}+1/2 } \times g(\kappa/\omega_{\rm in}) ~, 
\label{rapsp2}
\end{split}
\end{equation}
where $g(\kappa/\omega_{\rm in}) $ is a rather complicate function.\footnote{
It is given by 
\begin{equation}
\begin{split}
g(\kappa/\omega_{\rm in}) =  \frac{1}{\pi} \left ( \frac{ 2 \log \left (\frac{\kappa_{\rm eff} }{ \omega_{\rm in}}\right ) }{1 + (\omega_{\rm f} /\omega_{\rm in})^2}  -1\right )~,
\label{rapsp2pres}
\end{split}
\end{equation}
where $\kappa_{\rm eff}$ is the effective slope of the profile $h(\kappa t)$. It is given by
\begin{equation}
\begin{split}
\kappa_{\rm eff} &\doteq  \kappa \exp \left [{-  \gamma - \int dt dt'\left ( \partial_t h  \right ) \left ( \partial_{t'} h \right ) \log(\kappa \abs{t-t'}) } \right ] ~.
\label{keff}
\end{split} 
\end{equation}
where $\gamma$ is the Euler constant. The interest of these equations is that they apply for any $h(\kappa t)$ when $\kappa/\omega_{\rm in} \gg 1$. Moreover, $\kappa_{\rm eff}$ also governs the logarithmic growth of $\abs c$.}
We \newpage

\noindent numerically checked that Eqs.~\eqref{rapsp2} and~\eqref{rapsp2pres} apply for $n_{in} \lesssim 5$ and $\kappa \gtrsim 10 \omega_{\rm in}$. Hence, under these conditions, $\Delta n_r$ depends on $\kappa_{\rm eff}$ of \eq{keff}, but not on the exact shape of $h$ of Eq.~\eqref{eq:grad}. 

Second, for high occupation numbers and high $\kappa \Gamma / \omega_{\rm in}^2$, we observe in Fig.~\ref{fig:contourdeltanovern} a quadratic dependence in $\Gamma $. Lowering $\kappa \Gamma / \omega_{\rm in}^2$, we observe a transition from this quadratic behavior to a linear one. These numerical observations are in agreement with the analytical result 
\begin{equation}
\label{eq:deltan2}
\begin{split}
 \Delta n_r \sim  \left (\frac{\Gamma_{\rm in}}{\omega_{\rm in}}\right )^2 \frac{\omega_{\rm in}^2-\omega_{\rm f}^2}{\omega_{\rm in}^2+\omega_{\rm f}^2} + \mathcal{O}\left ( \frac{\Gamma_{\rm in}}{\kappa}\right )~, 
\end{split}
\end{equation}
which applies  in the limit $\Gamma_{\rm in} \ll \omega \ll \kappa \ll T$. 

In brief, Eq.~\eqref{rapsp2} and Eq.~\eqref{eq:deltan2} establish how $n^{out}$ of Eq.~\eqref{eq:nandcdiverge} converges towards the dispersive occupation number of Eq.~\eqref{outnc} when $\Gamma_{\rm in} / \omega_{\rm in} \ll 1$ and $\kappa \gg \omega_{\rm in}$. A similar analysis can be done for the coefficient $c^{out}$ of Eq.~\eqref{eq:nandcdiverge}, and gives similar results. 

\subsubsection{Final coherence level}

\begin{figure}
\includegraphics[width=1\linewidth]{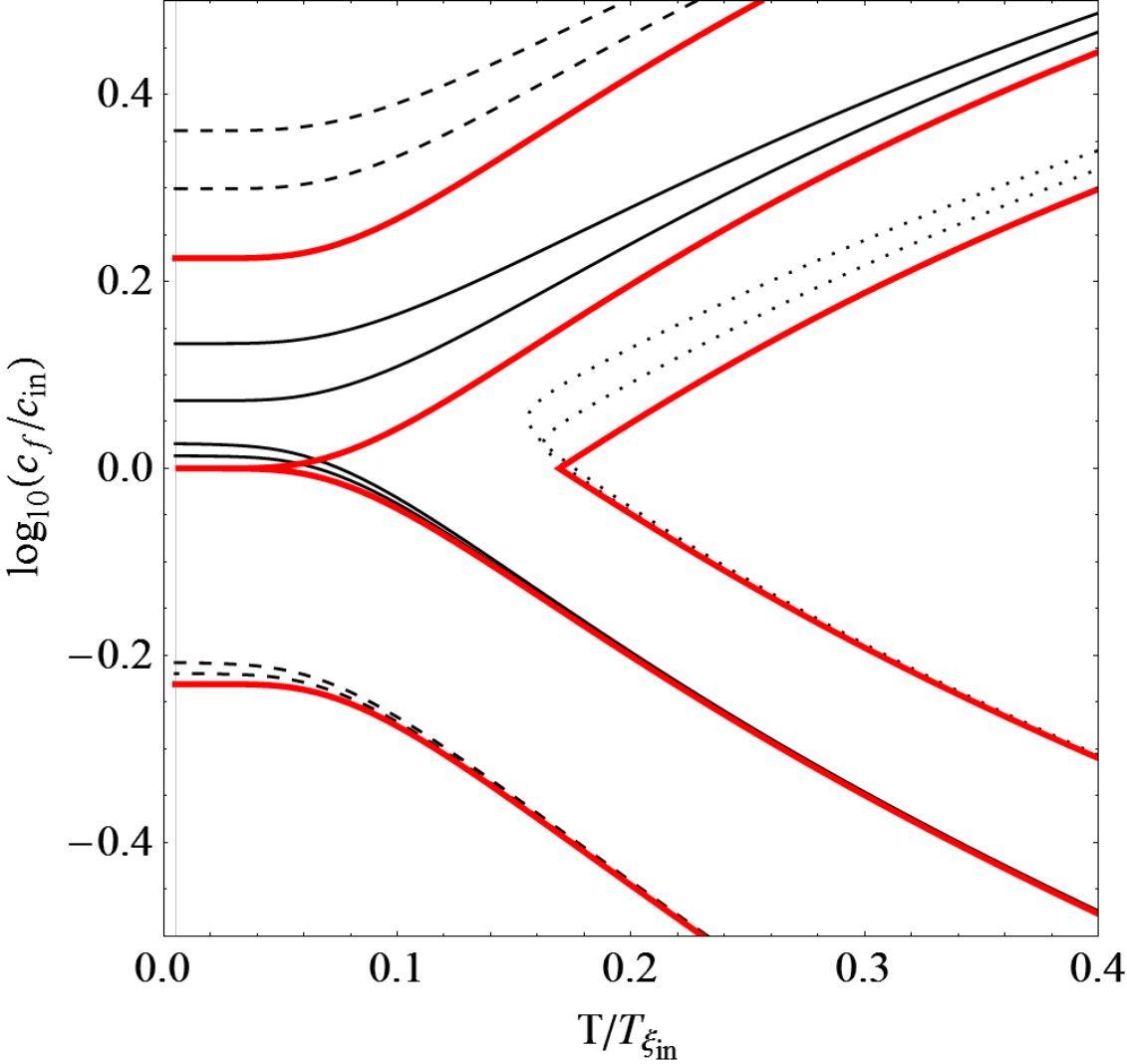}
\includegraphics[width=1\linewidth]{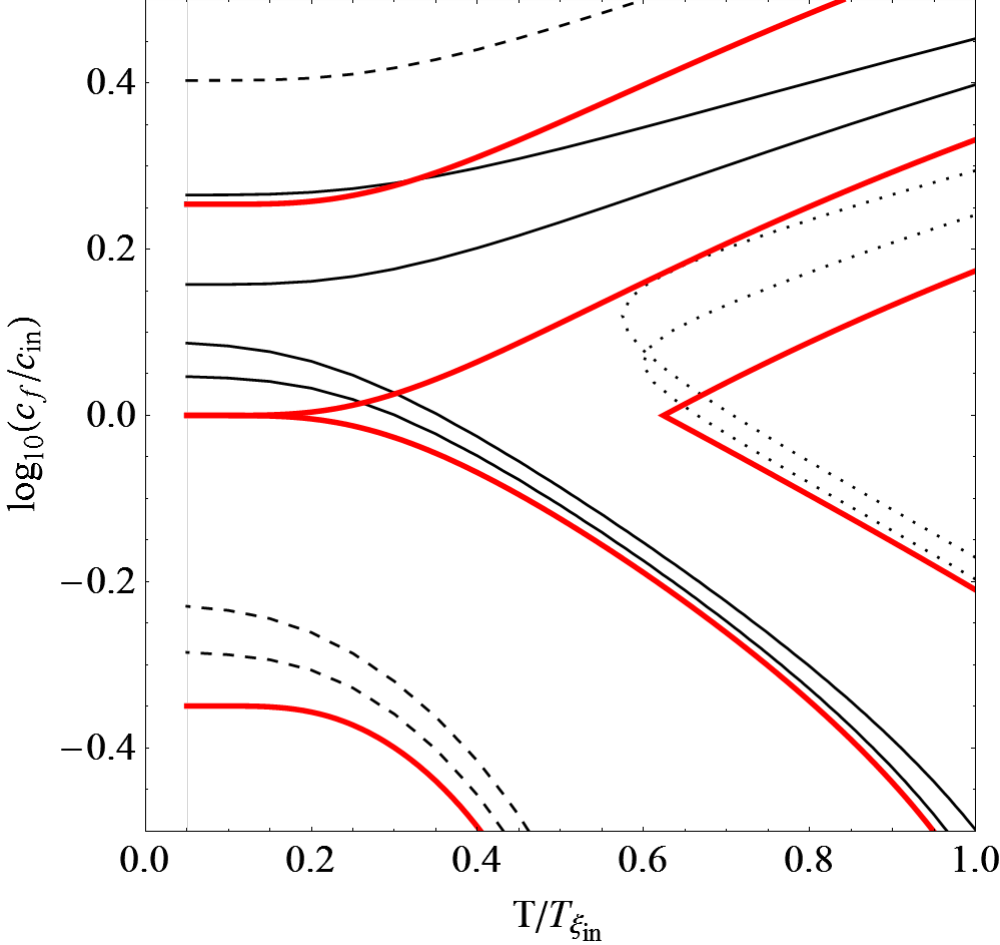} 
\caption{We represent the lines of $\Delta = -0.2$ (dashed), $0$(solid) and $0.2$ (dotted) in the plane $\{ T/T_{\xi_{\rm in}}, \log_{10} ( c_{\rm f} / c_{\rm in})  \}$, on the top panel for $k = 0.3 mc_{\rm in}$, and, on the bottom one for $k = mc_{\rm in}$. $\gamma_\bk^2$ takes $3$ values: from $0$, i.e. the dispersive case in thick red line, to ${0.5}$, and $\kappa = 10 m c_{\rm in}^2 $. For not too low temperatures, we observe that the value of $\Delta$ is robust when $c_{\rm f} < c_{\rm in}$ (which corresponds to an expanding universe), while it increases when $c_{\rm f} > c_{\rm in}$.} 
\label{fig:deltadissip}
\end{figure}

In fig.~\ref{fig:deltadissip}, we represent the coherence level $\Delta$ of Eq.~\eqref{deltak} as a function of the temperature and $c_{\rm f}/c_{\rm in}$, for two different values of $k$, namely $k/ m c_{\rm in}= 0.3$ and $1$, and three values of $\gamma^2$, namely $0, 0.25$ and $0.5$. The value of $\kappa$ is $\kappa = 10 m c_{\rm in}^2$. As one might have expected, we observe a continuous deviation of $\Delta$ for increasing values of the coupling $\gamma^2_\bk$. More surprisingly, we observe that increases of $c(t)$, $c_{\rm f} > c_{\rm in}$, and decreases $c_{\rm f} < c_{\rm in}$ behave very differently. In the first case, there is a large increase of $\Delta$, which implies a loss of coherence. On the contrary, in the second case, the value of $\Delta$ is robust, and some marginal gain of coherence can even be found. To validate these observations, we studied the behavior of $\Delta$ for different profiles of the smoothing function $h(\kappa t)$. Whenever $c_{\rm f} / c_{\rm in}$ not too close to 1, we obtained similar results, thereby showing that the choice of the profile of $h$ does not significantly matter. Instead, when $c_{\rm f} / c_{\rm in} \sim 1$, the behavior of $\Delta$ is less universal.

\begin{figure}
\includegraphics[width=1\linewidth]{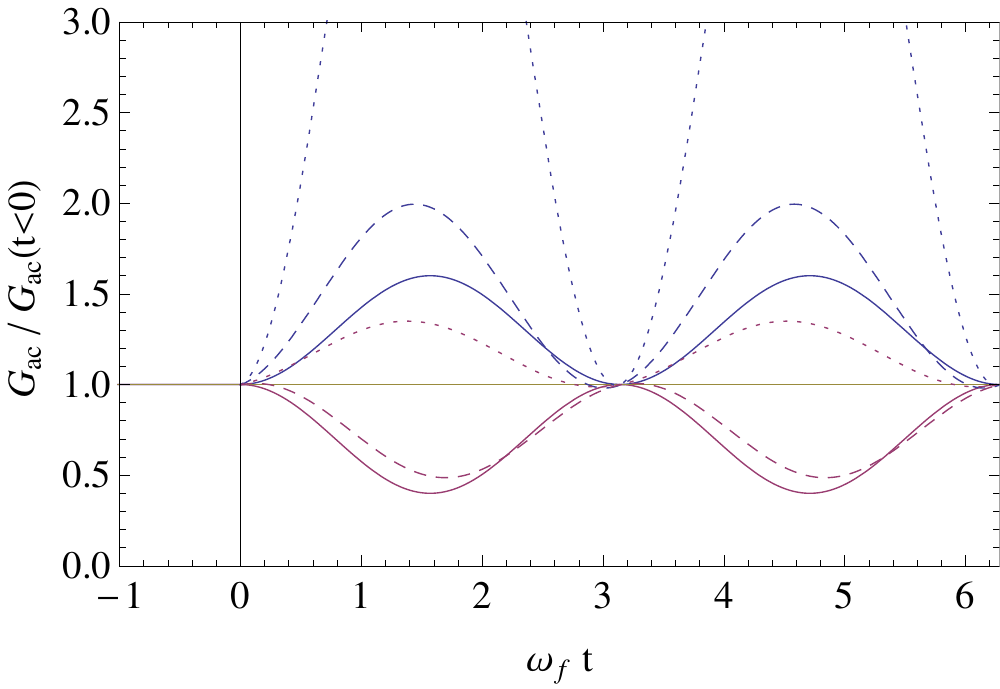}
\caption{We represent the anticommutator $G_{ac}$ normalized to its value before the jump, as a function of $\omega _{\rm f }t $, and for $T =T_{\xi_{\rm in}}$ and $k = 2 m c_{\rm in}$. The $3$ upper curves (in blue) represent an expanding universe ($c_{\rm f} = 0.5c_{\rm in}$), while the three lower (in purple) ones a contracting universe ($c_{\rm f} = 2c_{\rm in}$). In solid lines, one finds the two dispersive cases, in dashed lines, the dissipative cases with $\gamma^2=0.15$ and $\kappa = 100 m c_{\rm in }^2$, and in dotted lines, $\kappa$ has been increased to $10^4 m c_{\rm in }^2$ in order to see the logarithmic growth of $n$ and $c$ of \eq{outnc}. In all cases, $G_{ac}$ is $\mathcal{C}^{1}$ across the jump. Since the coherence is based on the minima of $G_{ac}$, the value of $\Delta$ is robust when $c_{\rm f}/c_{\rm in} < 1$, whereas it necessarily increases when $c_{\rm f}/c_{\rm in} > 1$. }
\label{fig:morerobust}
\end{figure}

In order to understand the different behaviors of $c_{\rm f} > c_{\rm in}$ and $c_{\rm f} < c_{\rm in}$, we represent in fig.~\ref{fig:morerobust} the anticommutator normalized to its value at the jump $G_{ac}(t=t') / G_{t=t'=0}$ as a function of $\omega_{\rm f} t$, and for different values of the steepness parameter $\kappa$. We first notice that this function is $\mathcal{C}^1$ across the jump, as were the modes in dispersive theories. This implies that one extremum of the anticommutator coincides with the value before the jump. In the absence of dissipation, one easily verifies that it is a minimum for $c_{\rm f} < c_{\rm in}$ and a maximum otherwise. For weak dissipation, by continuity in $\Gamma$, this must still be the case. Hence, when $c_{\rm f} < c_{\rm in}$, the minima of the anticommutator are fixed by the initial state. Instead, for $c_{\rm f} > c_{\rm in}$, they are fixed by the intensity of the jump and the injection of energy from the environment. As clearly seen in the Figure, this injection increases with $\kappa$ and explains why coherence is more robust when $c_{\rm f} < c_{\rm in}$ (i.e. in expanding universes).

\subsection{Constant dissipation rate}

\begin{figure*}
\begin{minipage}{0.45\linewidth}
\includegraphics[width=1\linewidth]{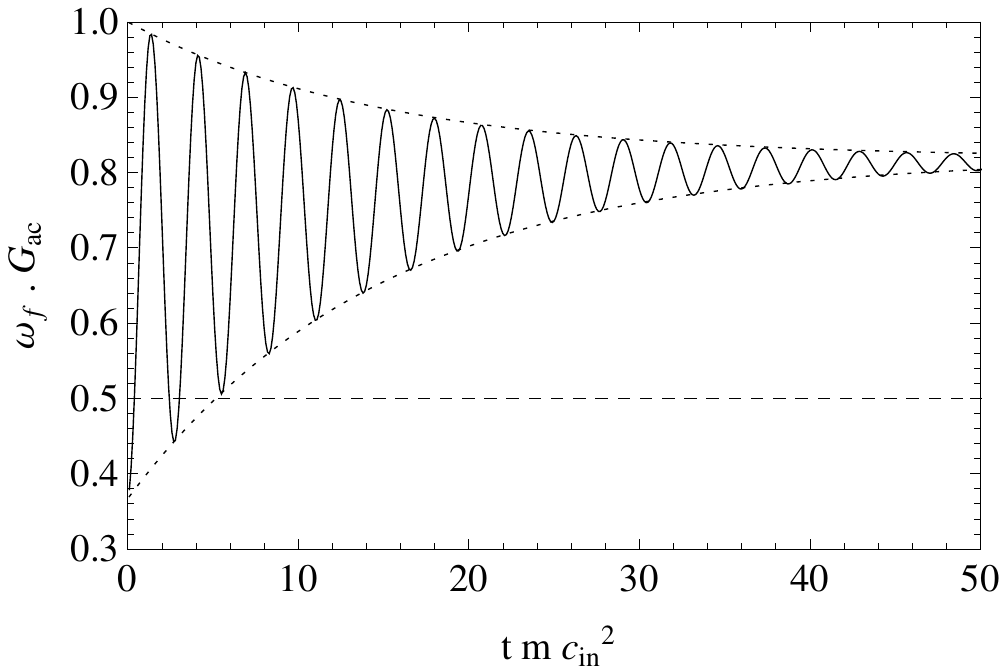}
\end{minipage}
\hspace{0.05\linewidth}
\begin{minipage}{0.45\linewidth}
\includegraphics[width=1\linewidth]{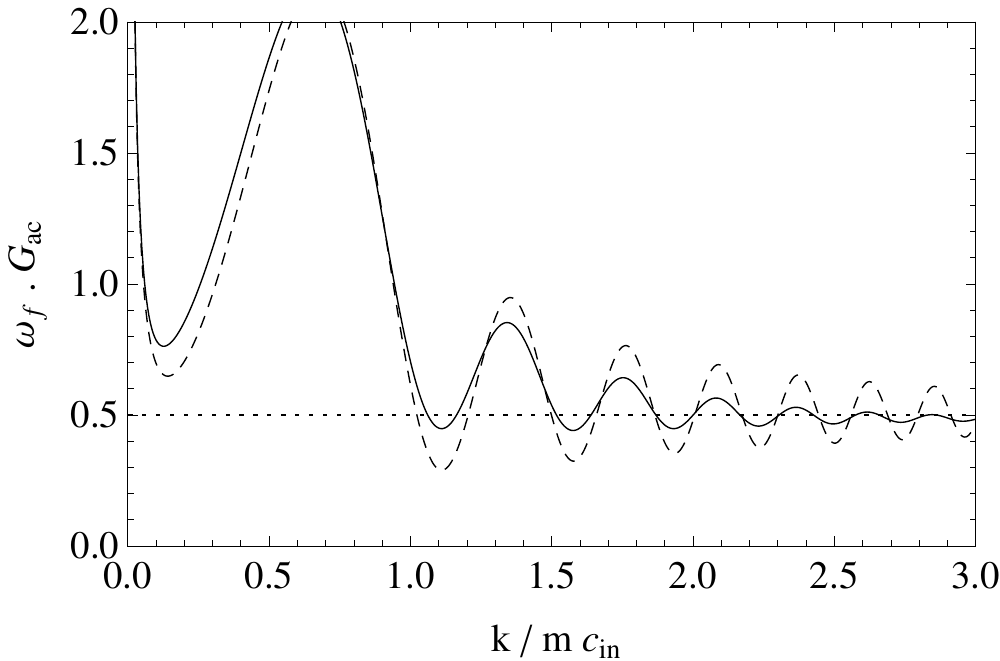}
\end{minipage}
\caption{We represent the product $\omega_{\rm f} \times G_{ac}(t,t'=t)$ where $G_{ac}$ is given in Eq.~\eqref{eq:gacofnc}, on the left panel, as a function of the adimensionalized time $t m c_{\rm in}^2$ for $k=1.5 m c_{\rm in}$, $\gamma_\bk^2 =0.03$ and $T = 0.8 T_{\xi_{\rm in}}$, and on the right panel, as a function of  $k/ m c_{\rm in}$ for $ t = 5/ m c_{\rm in}^2$, $\gamma_\bk^2 =0.05$ and $T = 0.5 T_{\xi_{\rm in}}$.  The dashed line on the right panel is the dissipative-less case  $\gamma_\bk^2 =0$. In all cases, the height of the jump is $c _{\rm f} / c_{\rm in} =0.1$.}
\label{fig:Gofk1}
\end{figure*}

In this section, we study our model when the dissipative rate $\Gamma$ is constant, as it is found for instance in polariton systems~\cite{Gerace:2012an,Koghee2013}. In this case, there is no unambiguous notion of ($out$) quanta, and this even though the anti-commutator of Eq.~\eqref{Gacdr} is well defined for all $t, t'$. Nevertheless, provided $\Gamma/\omega$ is low enough, we shall see that an approximate reading of the final state can be reached in term of the instantaneous particle representation based on \eq{chidec}. 

Because $\gamma$ is constant in \eq{eq:grad}, there is a simplification with respect to the previous subsection: no regularization is now needed since \eq{Gacdr} is finite. Moreover, the retarded Green function of \eq{Gret} is exactly known. It is given by
\begin{equation}
\label{Gretgrad}
\ep{-\Gamma (t-t')}  \times \left \{
\begin{array}{ll}
\theta(t-t')    \frac{\sin \omega (t-t')}{\omega}& \! \! \! \mbox{ for } t',t <0 \mbox{ or } t'>0~, \\
 \frac{\sin (\omega_{\rm f} t) \cos(\omega_{\rm in} t')}{\omega_{\rm f}}  - & \! \! \! \! \! \frac{\cos (\omega_{\rm f} t) \sin(\omega_{\rm in} t')}{ \omega_{\rm in} } \\
 &\hspace{-1cm}\mbox{ for } t' <0\mbox{ and } t >0~. \\
\end{array}\right .
\end{equation}
Hence, after the jump of $c$, for positive times, the Fourier transform of \eq{buildingblock} gives 
\begin{equation}
\label{eq:TFGret}
\begin{split}
\widetilde G_{\rm r}(t,\omega_\zeta)=&  \frac{\sqrt{  \Gamma} \ep{i\omega_\zeta t}}{\omega_{\rm f}^2 - ( \omega_\zeta- i \Gamma )^2} + \sqrt{\Gamma} \frac{ \ep{-\Gamma t+i\omega_{\rm f} t}}{2 \omega_{\rm f}}\\
\times &\bigg[\frac{ \omega_{\rm f} +( \omega_\zeta-i \Gamma) }{\omega_{\rm in}^2 - ( \omega_\zeta- i \Gamma)^2} - \frac{1 }{\omega_{\rm f} -  (\omega_\zeta - i\Gamma )} \bigg] \\
&+(\omega_{\rm f} \to - \omega_{\rm f}) ~.
\end{split} 
\end{equation}
This means that we (exactly) know the integrand of Eq.~\eqref{Gacdr}. The integral can be performed by analytical methods (by evaluating the residues of poles), and then recognizing the infinite sum as a finite sum of hypergeometric functions. The main results are presented below. 

\subsubsection{Two-point correlation function}

To discover the effects of dissipation, in Fig.~\ref{fig:Gofk1} we plot $\omega_{\rm f} \times G_{ac} (k,t=t') $ both as a function of time, as in Fig.~\ref{fig:Gacoft}, and as a function of the wave number $k$, as in Fig.~\ref{fig:Gacofk}. When considered as a function of $t$, we observe that the oscillations take place in a narrowing envelope. As expected, the latter follows an exponential decay in $e^{-2\Gamma t}$ towards the equilibrium value $\omega_{\rm f}\, G_{ac}^{eq} = \omega_{\rm f}\,   G_{ac}(t=t'\to \infty)$. This simple behavior implies that the non-separability of the state is quickly lost at high temperature. Indeed, a rough estimate of the lapse of time for the decoherence to happen is of the order $ (2 \Gamma n^{eq})^{-1}$, where $n^{eq}$ is the mean occupation number at equilibrium. Hence, when $n^{eq} \gg 1$, the time for the loss of coherence is smaller than the dissipative time $1/\Gamma$ by a factor $1/2 n^{eq}$. When considered as a function of $k$, on the right panel, we observe damped oscillations. For large $k$, they are more damped than those of the dispersive case (represented by a dashed line) since the decay rate $\Gamma \propto k^2$. 

\begin{figure}
\includegraphics[width=1\linewidth]{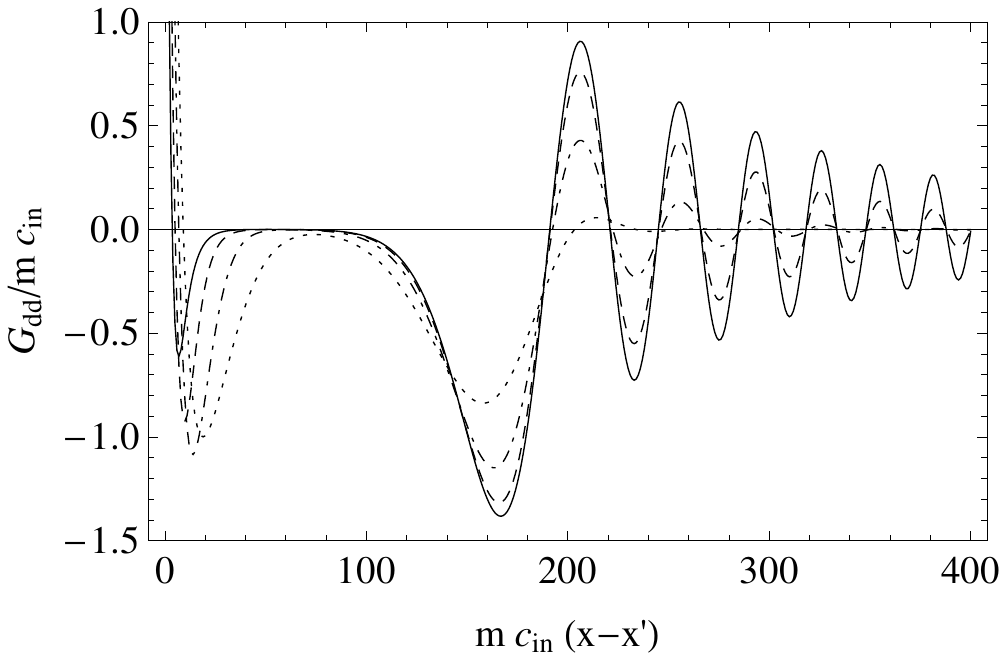} 
\caption{We represent the equal time density-density correlation of Eq.~\eqref{eq:defd} as a function of $ m c_{\rm in} \abs{x-x'}$, for $c_{\rm f}/c_{\rm in} =0.1$, and $T = T_{\xi_{\rm in}} $. We took $t = 7.5 /mc_{\rm f}^2$ and $4$ values of $\gamma_\bk^2$, namely $10^{-2}$ (solid), $10^{-1.5}$ (dashed), $10^{-1}$ (dot-dashed) and $10^{-0.5}$ (dotted). The parameter $n$ of Eq.~\eqref{gamma} is $n=0$. We observe a peak at $x=x'$ that is broadened by dissipation, and a series of peaks propagating away from the center with a group velocity higher than $c_{\rm f}$. The faster they propagate, the more damped they are since we have $\Gamma \sim k^2$. } 
\label{fig:deltaGofx}
\end{figure}

To further study the effects of increasing $\Gamma$, in Fig~\ref{fig:deltaGofx}, we represent the equal time density-density two-point function, see appendix~\ref{app:deltarhog1g2},
\begin{equation}
\label{eq:defd}
\begin{split}
G_{dd} &\doteq \frac{{\rm Tr}\left (\hat \rho_T \, \delta \hat \rho(t,x) \delta \hat \rho(t,x') \right ) 
}{\rho} \\
&= \int_{-\infty}^\infty \frac{d\bk}{\pi } \ep{ i \bk (x-x')}   c \xi k^2 G_{ac}^k(t,t'=t) ~,
\end{split}
\end{equation} 
for four values of $\gamma^2$, namely $\log_{10}(\gamma^2) =  - 2,\,  -1.5, \, -1$ and $-0.5$. We took a rather large $t = 7.5 /mc_{\rm f}^2$ to see the propagation of the phonon waves. As expected we observe a peak centered on $x=x'$ plus a series of peaks for $\abs{x-x'} > 2 c_{\rm f} t$. The first one is present even in vacuum, and is amplified by the mean occupation numbers $n_k > 0$. We see that it is broadened by dissipation. The series of peaks for $\abs{x-x'} > 2 c_{\rm f} t$ are due to the fact that the phonons are produced in pairs. As in inflationary cosmology~\cite{Campo:2003pa}, their amplitudes are fixed by the $c_k$ coefficients. These peaks propagate at different speed because of dispersion. The fastest are more damped because dispersion is anomalous, and because dissipation goes in $k^2$. We also notice that the first propagating peak is negative when working with $c_{\rm f} < c_{\rm in}$. (For $c_{\rm f} > c_{\rm in}$ instead, it would be positive.) We conclude by noticing that this plot gives no indication on whether the state is separable or not, mainly because $G_{dd}$ mixes different 2-mode sectors labeled by $k$, some of which being non-separable, but not all.

\subsubsection{Approximate particle interpretation and separability}
\label{sec:nbofpartdissip}

To interpret the properties of $G_{ac}$, we use the instantaneous particle representation based on $\chi^{dec}$ of Eq.~\eqref{chidec}. As already said, for interacting systems, there is no unique (canonical) way of defining the concept of particle. Hence the mean occupation number $n$ and the correlation term $c$ are somehow ambiguous. The issue is two fold and requires to treat separately the equilibrium and the out of equilibrium parts of $G_{ac}$. 

First, the out of equilibrium part $\delta G_{ac} \doteq G_{ac}^{dr} - G_{ac}^{eq} $ contains non oscillating terms which cannot be expressed as the anticommutator of $\chi^{dec}$ of Eq.~\eqref{chidec}. However, these terms decay as $\ep{- 2 \pi t  T}$, whereas oscillating terms decay as $\ep{- 2 \Gamma t }$. Hence, when $\Gamma < \pi  T$, they can be neglected for $t, t' \gg 1/(\pi T- \Gamma)$. When these conditions are fulfilled, one can define $\delta n (t_0)$ and $\delta c (t_0)$, the out of equilibrium value of the occupation number and the coherence at $t_0$, by
\begin{equation}
\label{eq:deltaGac}
\begin{split}
\delta G_{ac}(t,t') & \sim   \delta n (t_0)\varphi^{dec} (t;t_0) [\varphi^{dec} (t';t_0)]^* \\
&+ \delta c(t_0) \varphi^{dec} (t;t_0) \varphi^{dec} (t';t_0) +cc~,
\end{split}
\end{equation}
where $\varphi^{dec} (t;t_0) = \ep{- \Gamma (t-t_0)} \ep{ - i\omega_{\rm f} (t-t_0)} / \sqrt{2 \omega_{\rm f}}$ is the decaying solution of Eq.~\eqref{chidec} which contains only positive frequency and which is unit Wronskian at $t=t_0$. Since $\delta G_{ac}(t,t')$ is independent of $t_0$, one immediately deduces that
\begin{equation}
\delta n (t_0) = \delta n(0) \ep{- 2 \Gamma t_0} ~,\quad \vert \delta c(t_0) \vert = 
\vert \delta c(0) \vert \ep{- 2 \Gamma t_0} ~.
\end{equation}
This matches the behavior of the envelope observed in Fig.~\ref{fig:Gofk1}. In the limit of small dissipation, one finds that the initial values obey
\begin{equation}
\begin{split}
\delta n(0) &= \delta n^{disp} + \mathcal{O} \left (\frac{\Gamma}{ \omega_{\rm in}}+\frac{\Gamma}{ T}   \right )~,\\
\delta c(0) &= c^{disp}+  \mathcal{O} \left (\frac{\Gamma}{ \omega_{\rm in}}+\frac{\Gamma}{  T}   \right )~,
\end{split}
\end{equation}
where $\delta n^{disp} = n^{out} - n^{eq}$ and $c^{disp}=c^{out}$ are the corresponding quantities evaluated with the dispersive case $\gamma = 0$. The values of $n^{out}$ and $c^{out}$ are given in sec.~\ref{sec:nodiss}, and $n^{eq}$ is the mean occupation number in a thermal bath.

Second, a similar analysis of the equilibrium part of $G_{ac}$ gives
\begin{equation}
\label{eq:eqGacoftt}
\begin{split}
G_{ac}^{eq}(t,t') & = \ep{-\Gamma\abs{t-t'}}\, G_{ac}^{th,disp}(t,t') + {\mathcal O} (\frac{\Gamma}{\omega_{\rm f}})~,
\end{split}
\end{equation}
where $G_{ac}^{th,disp}$ is the corresponding dispersive anticommutator in a thermal state. It is worth noticing that in the presence of dissipation, the rescaled anticommutator can be smaller than $1/2$, as can be seen in Fig.~\ref{fig:Gofk1} for high $k$. This is because the $\phi$ field is still interacting with the environment. Yet, in the limit $\Gamma \abs{t-t'} \ll 1$ and $\Gamma/ \omega_{\rm f} \ll 1$, $G_{ac}^{eq}$ is indistinguishable from $G_{ac}^{th,disp}$. We can then work with $2 n^{eq}+1 = { \coth[ \omega_{\rm f} /2T  ]} $. Doing so, we get an imprecision of the order of $\Gamma / \omega_{\rm f}$. 

\begin{figure}
\includegraphics[width=1\linewidth]{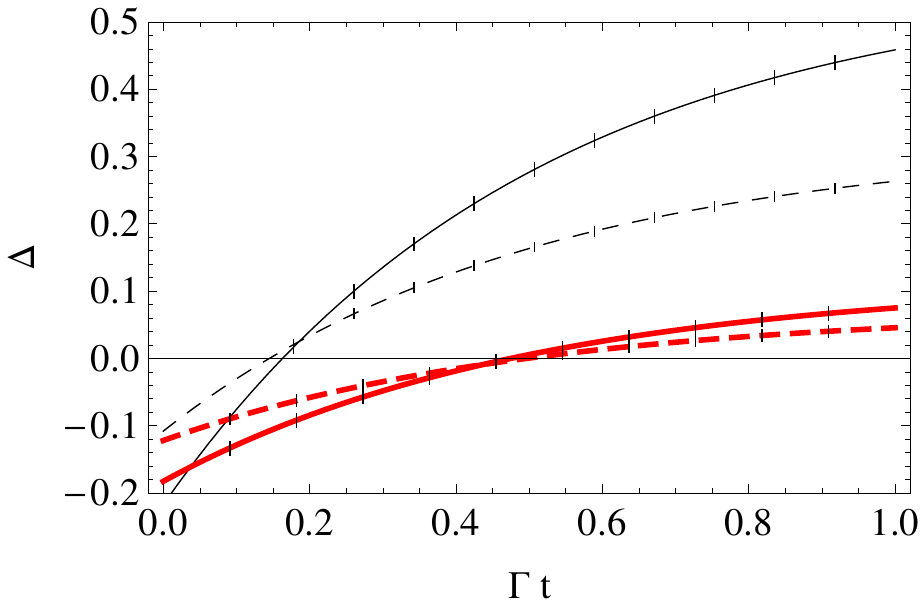} 
\caption{ We represent the coherence level $\Delta$ as a function of $\Gamma t$, 
for $T= T_{\xi_{\rm in}}/2$ and $\gamma_\bk ^2 =0.01$. We consider two values of $c_{\rm f} / c_{\rm in} = 0.1$ (solid) or $0.5$ (dashed), and two values of $k / m c_{\rm in} = 1$ (black) or $1.5$ (thick red). The imprecision in the value of $\Delta$ is indicated by vertical bars. In the present weakly dissipative cases, the spread of $\Delta$ is of the order $0.02$. Therefore, the moment where the non-separability of the state is lost is known with some precision. }
\label{fig:deltadissipoft}
\end{figure}

Having identified $n = \delta n + n^{eq}$ and $ c$, we can compute the coherence level $\Delta$ of Eq.~\eqref{deltak}, which of course inherits the imprecision coming from $n^{eq}$. In fig.~\ref{fig:deltadissipoft}, we represent $\Delta_\bk$ and its imprecision as a function of time for four different cases. As already discussed, we notice that the non-separability of the state is lost in a time much smaller than $1/\Gamma$. We also notice that when $\Gamma/\omega$ is low enough, the imprecision in $\Delta$ does not significantly affect our ability to predict when non-separability will be lost. In brief, for low values of $\Gamma/\omega$ and $\Gamma/T$, the anti-commutator $G_{ac}$ can be reliably interpreted at any time $t_0$ using the instantaneous particle representation based on $\chi^{dec}(t,t_0)$ of Eq.~\eqref{chidec}.

\section{Conclusions}
\label{sec:conclu}

In this paper, we computed the spectral properties $n_k$ and the coherence coefficient $c_k$ of quasi-particles produced when a sudden change is applied to a one-dimensional homogeneous system. We took into account both the effects of an initial temperature, and the fact that the quasi-particles are coupled to a reservoir of modes, something which induces dissipative effects. For definiteness the quasi-particles are taken to be Bogoliubov phonons propagating in an elongated atomic condensate. Yet our results should apply, mutatis mutandis, to all weakly dissipative systems. We are currently extending our treatment to polariton fluids.

For simplicity, we worked with a quadratic action. Importantly, this allows us to compute the anticommutator for non-equal times, see Eqs.~\eqref{Gact0} to~\eqref{Gacmix}, something which not generally done when using the truncated Wigner method~\cite{0953-4075-35-17-301}, but which could be very useful for future experiments. Because our system is coupled to a bath, $n_k$ and $c_k$ are {\it a posteriori} extracted from the anti-commutator of the phonon field, see Eq.~\eqref{eq:nandcdiverge} and~\eqref{eq:deltaGac}. Then, to distinguish classical correlations from quantum entanglement, we used the fact that negative values of the parameter $\Delta_k$ of Eq.~\eqref{deltak} corresponds to non-separable states (for the Gaussian states we consider).

When neglecting dissipative effects, we studied the competition between the squeezing of the quasi-particles state, which is induced by the sudden change, and the initial temperature which increases the contribution of stimulated effects, see Eq.~\eqref{deltaout}. In Fig.~\ref{fig:Gacoft}, one clearly sees that the value of the minima of the equal time anticommutator allows to know if the state is separable or not. The outcome of the competition is summarized in Fig.~\ref{deltadispgradino1}, where the coherence parameter $\Delta$ is plotted as a function of the sudden change of the sound speed and the temperature. We applied our analysis to the recent experiment of Ref.~\cite{PhysRevLett.109.220401} and concluded that one should either increase the change of $c$, or work with a lower temperature in order to obtain a non-separable state.

We then included dissipative effects. When there is no (significant) dissipation after the sudden change, we showed in Fig.~\ref{fig:contourdeltanovern} how the final number of particles is progressively affected by increasing dissipation. When the initial occupation number is low, the deviations are linear in the decay rate $\Gamma$, whereas they are quadratic for occupation numbers $n^{\rm in } \gtrsim 5$. Interestingly, we observed in Fig.~\ref{fig:deltadissip} that dissipation, on the one hand,  hardly affects the coherence parameter $\Delta$ when the sudden change is due to a decrease of the sound speed (something which corresponds to an expanding universe when using the analogy with gravity), and on the other hand severely reduces the coherence when the sound speed increases. This discrepancy is further studied in Fig.~\ref{fig:morerobust} which illustrates the key role played by the ${\cal C}^1$ character of the anticommutator across the sudden change. 

We also studied the case when the dissipative rate is constant. In this case, the main effect of dissipation on the anticommutator is the expected damping towards the equilibrium value, see Fig.~\ref{fig:Gofk1}. As a result, for high occupation numbers, the non-separability of the state is lost in a time much shorter than the inverse decay rate, see Fig.~\ref{fig:deltadissipoft}. In spite of the fact that the quasi-particles are still coupled to the environment, we showed that a reliable study of this loss can be performed for weakly dissipative systems, i.e. with $\Gamma/\omega \ll 1$. On the contrary, for strongly interacting systems, i.e. rapidly decaying quasi-particles, we believe the notion of non-separability cannot be meaningfully implemented.

\acknowledgments
We thank Scott Robertson for a careful reading of our manuscript, and Iacopo Carusotto for interesting discussions. 
This work, which is part of the project QEAGE (Quantum Effects in Analogue Gravity Experiments), was supported by the French National Research Agency under the Program Investing in the Future Grant n°ANR-11-IDEX-0003-02.

\appendix

\section{Action for relative density fluctuations}
\label{app:condensation}

We briefly review how  Eq.~\eqref{SPhi} is obtained from the action of the atomic field, for more details see~\cite{Macher:2009nz}. We then explain how Eq.~\eqref{Sint} can also be derived from an action defined at the atomic level (rather than that of the quasi-particles). 

The action for the second quantized field describing a dilute ultracold atomic gas is~\cite{Dalfovo:1999zz}
\begin{equation}
\begin{split}
S  = \int dt dx \bigg[ i \hat \Phi^\dagger \partial_t \hat \Phi &- \frac{1}{2m} \partial_x \hat \Phi^\dagger \partial_x \hat \Phi - V \hat \Phi^\dagger \hat \Phi \\
&- \frac{g_{at}}{2} \hat \Phi^\dagger \hat \Phi^\dagger \hat \Phi \hat \Phi\bigg]~. 
\end{split}
\end{equation}
One decomposes $\hat \Phi = \Phi_{\rm cond} ( 1+ \hat \phi)$, where $\Phi_{\rm cond}(t,x)$ is the mean field describing the condensed atoms, solution of the Gross Pitaevskii equation
\begin{equation}
\label{grosspit}
\begin{split}
i\partial_t \Phi_{\rm cond} =   \left ( -\frac{1}{2m} \partial_x^2  + V + g_{at}\Phi_{\rm cond}^* \Phi_{\rm cond} \right )\Phi_{\rm cond}  ~,
\end{split}
\end{equation}
and where $\hat \phi $ describes {relative} density fluctuations. The action is then expanded in powers of $\hat \phi$. Using Eq.~\eqref{grosspit}, the quadratic part is 
\begin{equation}
\label{SPhi1}
\begin{split}
S_{\phi} = \int dt dx \rho \bigg[& i \hat \phi^\dagger  (\partial_t + v \partial_x) \hat \phi -  \frac{1}{2m} \partial_x (\hat \phi^\dagger )   \partial_x \hat \phi \\
& - \frac{m c^2}{2} (\hat \phi^2 + (\hat \phi^\dagger )^2 +2 \hat \phi^\dagger \hat \phi )\bigg]~,\\
\end{split}
\end{equation}
where  $\rho \doteq \vert \Phi_{\rm cond} \vert^2 $, $ 2i m  v \doteq \partial_x ( \ln \Phi_{\rm cond} - \ln \Phi_{\rm cond}^*)$ and $c^2 \doteq g_{at} \rho/m$ are arbitrary functions of $t$ and $x$ obeying $\partial_t \rho + \partial_x(\rho v) = 0$. When considering homogeneous condensates at rest ($v=0$), one obtains Eq.~\eqref{SPhi} with $\rho = cst.$

One can also derive Eq.~\eqref{Sint} when coupling the environment field $\Psi_\zeta$ to the atomic field. More precisely, to get Eq.~\eqref{Sint}, one should work with
\begin{equation}
\begin{split}
S_{\it int}\!  &=\!  - \! \!  \int \! \!  dt dx   \left \{ [(\hat \Phi^{\dagger})^{\alpha}  \hat \Phi^\alpha] \,\tilde g ( \partial_x)  \partial_t (\int \! \! d\zeta \bar \Psi_\zeta) \right \}~.
\label{Sint2}
\end{split}
\end{equation}
Then, using Eq.~\eqref{SPsi}, the equations of motion are
\begin{subequations}
\label{eq:eomincondfield}
\begin{align}
\label{eq:eomincondfieldPhi}
i\partial_t \hat \Phi &=   \left ( -\frac{1}{2m} \partial_x^2  + V + g_{at}\hat \Phi^\dagger \hat \Phi \right )\hat \Phi \\
\nonumber &\hspace{1cm}+ \alpha (\hat \Phi^{\dagger})^{\alpha-1}  \hat \Phi^{\alpha}  \, \tilde g(\partial_x) \partial_t (\int \! \! d\zeta \bar \Psi_\zeta)~,\\
(\partial_t^2 &+ \omega_\zeta^2) \bar \Psi_\zeta = \partial_t \tilde g ( -\partial_x) [(\hat \Phi^{\dagger})^{\alpha}  \hat \Phi^\alpha]~.
\end{align}
\end{subequations}
Hence, $\bar \Psi_\zeta$ will also condense. We can write it 
\begin{equation}
\begin{split}
\bar \Psi_\zeta = \Psi_\zeta^{ \rm cond} + \hat \Psi_\zeta~.
\end{split}
\end{equation}
From \eq{eq:eomincondfieldPhi}, we get  a modified Gross-Pitaevskii equation
\begin{equation}
\begin{split}
i\partial_t \Phi_{\rm cond} =  \mathcal{H}_{{\rm cond}} \Phi_{\rm cond} ~,
\end{split}
\end{equation}
with 
\begin{equation}
\begin{split}
\mathcal{H}_{{\rm cond}} = \frac{-\partial_x^2}{2m}   &+  V + g_{at}\abs{\Phi_{\rm cond}}^2 \\
& +  \alpha \abs{\Phi_{\rm cond}}^{\alpha-1}   \, \tilde g(\partial_x) \partial_t (\int \! \! d\zeta \bar \Psi_\zeta^{ \rm cond} ) ~.
\end{split}
\end{equation}
Since this last term does not contain any operator acting on $\Phi_{\rm cond}$, the conservation equation $\partial_t \rho + \partial_x (\rho v)=0$ is still valid. Moreover, it gives rise, in Eq.~\eqref{SPhi1}, to a term 
\begin{equation}
\label{eq:addedterm}
\begin{split}
\delta S_{\phi } = \int dt dx \alpha  \phi^\dagger \phi \rho^{\alpha}   \, \tilde g(\partial_x) \partial_t (\int \! \! d\zeta \Psi_\zeta^{ \rm cond} )~.
\end{split}
\end{equation}
On the other hand, Eq.~\eqref{Sint2} gives two contributions given by
\begin{equation}
\begin{split}
S_{int}^{(1)} &=   -  \! \! \int    \! \!dt dx    \bigg \{ [ \frac{\alpha(\alpha -1)}{2}(\hat \phi^{\dagger})^{2} + \alpha^2 \hat \phi^{\dagger} \hat \phi +  \frac{\alpha(\alpha -1)}{2} \hat \phi^2]\\
&\hspace{2cm}\times \rho^\alpha \,\tilde g ( \partial_x)  \partial_t (\int \! \! d\zeta  \Psi_\zeta^{ \rm cond} ) \bigg \}~,\\
S_{int}^{(2)} &=  -  \int   \! \! dt dx  \left \{ \alpha \rho^\alpha[\hat \phi^{\dagger}+  \hat \phi] \,\tilde g ( \partial_x)  \partial_t (\int \! \! d\zeta \hat \Psi_\zeta) \right \}~.
\end{split}
\end{equation}
Since the first one is second order in $\phi$ and zero order in $ \hat \Psi_\zeta$, it combines with Eq.~\eqref{eq:addedterm} to give 
\begin{equation}
 \begin{split}
\delta S_{\phi }+ S_{int}^{(1)} =\! \!\int \! \!dt dx &\frac{\alpha(\alpha-1)}{2}  (\phi^\dagger + \phi)^2 \rho^{\alpha} \\
&\times  \tilde g(\partial_x) \partial_t (\int \! \! d\zeta \Psi_\zeta^{ \rm cond} )  ~.
 \end{split}
 \end{equation} 
Its role is to modify the speed of sound which is now given by 
\begin{equation}
\begin{split}
m c^2 \doteq g_{at} \rho + \frac{\alpha(\alpha-1)}{2}   \rho^{\alpha-1}  \, \tilde g(\partial_x) \partial_t (\int \! \! d\zeta \Psi_\zeta^{ \rm cond} )~. 
\end{split}
\end{equation}
The second contribution $S_{int}^{(2)}$ gives rise to Eq.~\eqref{Sint}, with $ \alpha \tilde{g}(-\partial_x) = g  \xi^{\alpha+1/2} (\xi \partial_x)^n$. Hence, we have demonstrated that our model based on Eqs.~\eqref{SPhi} and~\eqref{Sint} can be derived from actions involving only the atomic field $\hat \Phi$. 

Notice that in a polariton system~\cite{Gerace:2012an,Koghee2013}, the dissipative processes occur at the level of the number of photons, and not only at the level of the quasi-particles (phonons of the photon fluid). This means that the coupling between $\Psi$ and $\Phi_{\rm pola}$ should be of the form
\begin{equation}
\begin{split}
S_{\it int, pola}\!  &=\!  - \! \!  \int \! \!  dt dx  \left \{ \gamma \hat \Phi_{\rm pola} \, \partial_t (\int \! \! d\zeta \hat \Psi_\zeta) + h.c. \right \}~. 
\label{Sintpola}
\end{split}
\end{equation}
We are currently studying this case. 

\section{On other observables}
\label{app:deltarhog1g2}

We establish the dictionary between the $\chi$ field of Eq.~\eqref{defchi} and its anticommutator $G_{ac}$ of Eq.~\eqref{Gact0}, and two other languages often used in the literature, namely, on the one hand, the phase and density fluctuations $\theta$ and $\delta \rho$, and, on the other hand, the so called $g_1$ and $g_2$ functions~\cite{PhysRevA.67.053615,Dalfovo:1999zz}. These functions are expressed in terms of the atomic field $\hat \Phi$ as 
\begin{equation}
\label{deltarhog1g2def}
\begin{split}
\hat \Phi(x,t) &= \Phi_{cond} \left ( 1+ \frac{\delta \hat \rho }{ 2 \rho }   \right ) \ep{ i \hat \theta}~,\\
g_1 (x,t,x',t') \! &\doteq \! {\rm Tr}\left (\hat \rho_T \,: \{ \hat \Phi^\dagger(x,t) ,  \hat \Phi(x',t')\} : \right )  ~,\\
g_2 (x,t,x',t') \! &\doteq \!   \frac{{\rm Tr} \left (\hat \rho_T \, \hat \Phi^\dagger(x,t) \hat \Phi^\dagger(x',t') \hat  \Phi(x',t') \hat \Phi(x,t) \right )}{g_1 (x,t,x,t) g_1 (x',t',x',t')}~,
\end{split}
\end{equation}
where $: \ :$ means normal ordered in the $\Phi$ field operator. In homogeneous condensates, in momentum space, and to linear order, we have
\begin{equation}
\begin{split}
\hat \chi_\bk &=  - \frac{\delta \hat \rho_\bk}{\sqrt{2 \rho c \xi k^2}} ~,\quad \partial_t \hat \chi_\bk =  \sqrt{2 \rho\xi  c  k^2 } \hat  \theta_\bk~. 
\end{split}
\end{equation}
Hence, the three anti-commutators are related to our $G_{ac}$ by 
\begin{equation}
\begin{split}
{\rm Tr}\left (\hat \rho_T \, \delta \hat \rho(t) \delta \hat \rho(t') \right ) &=  2 \rho c \xi k^2 G_{ac}^k(t,t')  ~,\\
{\rm Tr}\left (\hat \rho_T \, \hat \theta(t) \delta \hat \rho (t')\right ) &= - \partial_t G_{ac}^k(t,t') ~,\\
{\rm Tr}\left (\hat \rho_T \, \hat \theta(t) \hat \theta(t') \right ) &=   \frac{ \partial_t \partial_{t'} G_{ac}^k(t,t') }{2 \rho c \xi k^2}  ~. 
\end{split}
\end{equation}
The knowledge of $G_{ac}$ of Eq.~\eqref{Gact0} thus fixes the three of them.

Similarly, for $\bk \neq 0$, we have
\begin{equation}
\begin{split}
g_1^k (t,t') = & \left [ \frac{c\xi k^2}{2} + \frac{1}{2c\xi k^2} \partial_t \partial_{t'} \right ]G_{ac}^k(t,t') \\
&+ i \partial_t G_c^k(t,t')~, 
\end{split}
\end{equation}
where $ G_c^k(t,t')$ is the commutator of $\chi$ (which is imaginary), and
\begin{equation}
\begin{split}
g_2^k (t,t') &= \frac{2}{\rho} \bigg[ c\xi k^2 G_{ac} ^k(t,t') - i \partial_t G_c^k(t,t')  \bigg]~.
\end{split}
\end{equation}
Using Eq.~\eqref{eq:gacofnc}, we see that a measurement of $g_1(t,t'=t)$, or $g_2(t,t'=t)$, for various $t$ is sufficient to extract $n$ and $c$, and therefore to distinguish non-separable states from separable ones. For instance, the minimum value of $g_2(t,t'=t)$ is proportional to $ n - \abs{c} + (1- \omega_{\rm f} / c\xi k^2)/2$. Hence, when knowing $\omega_{\rm f} / c\xi k^2$ and $\rho$, measuring the minima of $g_2(t,t'=t)$ is sufficient to distinguish non separable states. 

\section{Approximating Eq.~\texorpdfstring{\eqref{eq:defR}}{44}}
\label{app:renorm2}

The goal of this appendix is to approximatively evaluate Eq.~\eqref{eq:defR}. To do so, we first consider that $h$ is constant for negative times.\footnote{This simplifies the study, but is not necessary. When it is not the case, we should apply the same treatment to positive and negative times.} 
Hence, the part of the integral that runs on negative times is easy to handle and gives
\begin{equation}
\label{eq:negtimeR}
\begin{split}
&\int_{-\infty}^0 d\tau h(\kappa \tau ) \ep{i \omega_{\zeta} \tau } \ep{ - \frac{\Gamma_{\rm in}}{\kappa} \int_{\kappa \tau}^\infty h^2(z) dz } \sqrt{2 \omega_{\rm f}} \varphi^{out}( \tau) \\
&\sim \ep{ - \frac{\Gamma_{\rm in}}{\kappa} \int_{0}^\infty h^2 }  \frac{ \omega_{\rm f} +( \omega_\zeta-i \Gamma_{\rm in}) }{\omega_{\rm in}^2 - ( \omega_\zeta- i \Gamma_{\rm in} )^2}~.
\end{split}
\end{equation}
Here, we neglected the effect of the change of $\omega$ for positive times on $\varphi^{out}$. The deviation is generically of order $\Gamma_{\rm in}^2 / \omega_{\rm f} \kappa $.\footnote{
Note that for low values of $\kappa$, a WKB approximation gives the same first term, with an upper bound on the approximation given by $\Gamma^2/\omega^2 \left (1+ \kappa / \omega\right )$.}
To get this bound, we can write, for positive times, $\varphi^{out} = \ep{- i \omega_{\rm f}\tau}/ \sqrt{2\omega_{\rm f}} + \varphi_1$, and perturbatively in $\varphi_1$, get to $\abs{\varphi_1}  \sqrt{2\omega_{\rm f}} < \Gamma_{\rm in}^2 / \omega_{\rm f} \kappa  \int_0^\infty f^2  $. This bound can be relaxed order by order by solving Eq.~\eqref{eq:homoeqvarphi} with a source. This is not the goal of this appendix. For positive times, we now make the same approximation and get
\begin{equation}
\label{eq:postimeR}
\begin{split}
\int_{0}^\infty d\tau h(\kappa \tau ) \ep{i \omega_{\zeta} \tau } \ep{ - \frac{\Gamma_{\rm in}}{\kappa} \int_{\kappa \tau}^\infty h^2(z) dz } \sqrt{2 \omega_{\rm f}} \varphi^{out}( \tau) \\
\sim   \int_{0}^\infty d\tau h(\tau) \ep{ - \frac{\Gamma_{\rm in}}{\kappa} \int_{\tau}^\infty h^2 } \ep{i (\omega_{\zeta}- \omega_{\rm f}) \tau }~.
\end{split}
\end{equation}
We can now compute this last integral perturbatively in $\Gamma / \kappa$. To get coherent results and to get rid of the $1/ \omega_{\zeta}$ term, the same expansion is necessary in Eq.~\eqref{eq:negtimeR}.

Since for cases we consider (i.e., sudden change, i.e., $\Gamma \ll \omega \ll \kappa$), $ \Gamma_{\rm in} / \kappa > \Gamma_{\rm in}^2 / \omega_{\rm f} \kappa  > \Gamma_{\rm in}^2 / \kappa^2 $, the expansion in $\Gamma_{\rm in} / \kappa$ should be done to first order maximum. To this order, $R$ becomes
\begin{equation}
\label{eq:Rforanyf}
\begin{split}
R&= \bigg[\left (1 { - \frac{\Gamma_{\rm in}}{\kappa}  \int_{0}^\infty h^2 } \right ) \frac{ \omega_{\rm f} +( \omega_\zeta-i \Gamma_{\rm in}) }{\omega_{\rm in}^2 - ( \omega_\zeta- i \Gamma_{\rm in} )^2}\\
&+\int_{0}^\infty d\tau  h(\kappa \tau) \left(1  - \frac{\Gamma_{\rm in}}{\kappa} \int_{\tau}^\infty h^2 \right ) \ep{i (\omega_{\zeta}- \omega_{\rm f}) \tau } \bigg]\\
&\hspace{1cm} \times  \left [1 + \mathcal{O} \left ( \frac{\Gamma_{\rm in}^2 }{ \omega_{\rm f} \kappa }\right )\right ]~.
\end{split}
\end{equation}
When working with $h$ of Eq.~\eqref{eq:chosenf}, one obtains 
\begin{equation}
\label{eq:Rforchosenf}
\begin{split}
R&\sim  \frac{ \omega_{\rm f} +( \omega_\zeta-i \Gamma_{\rm in}) }{\omega_{\rm in}^2 - ( \omega_\zeta- i \Gamma_{\rm in} )^2} + \frac{ h_1\left (\frac{i (\omega_\zeta -\omega_{\rm f})}{\kappa }\right ) }{\omega_\zeta -\omega_{\rm f}} \\
 &- \frac{\Gamma_{\rm in}}{3 \kappa} \left  [ \frac{ \omega_{\rm f} +( \omega_\zeta-i \Gamma_{\rm in}) }{\omega_{\rm in}^2 - ( \omega_\zeta- i \Gamma_{\rm in} )^2}+\frac{ h_4\left (\frac{i (\omega_\zeta -\omega_{\rm f})}{\kappa }\right )}{\omega_\zeta -\omega_{\rm f}} \right ] ~.
\end{split}
\end{equation}
where $ \ep{x} = \sum_{k=0}^{n} x^k/k! -  h_n (x) x^n/n!  $ defines the remainder term of order $n$, $h_n(x)$, of the Taylor expansion of the exponential function. We notice that $h_n(x)\propto x$ for $x\to 0$ so that $R$ is regular at $\omega_\zeta = \omega_{\rm f}$. Moreover, at large $x$, $h_n (x) \sim   \ep{x} n! /x^n + ( 1+ \mathcal{O}(1/x))$ so that for purely $R \sim 1/ \omega_\zeta^2$ at large $\omega_\zeta$.

To complete the study, we checked the validity of \eq{eq:Rforchosenf} by numerically evaluating $R$. When considering only first line of Eq.~\eqref{eq:Rforchosenf}, we observed that the relative error is smaller than $\Gamma / \kappa $, as predicted. When including the second line, the relative error remains smaller than $\Gamma^2 / \kappa \omega_{\rm f}$.

\bibliographystyle{../biblio/h-physrev}
\bibliography{../biblio/bibliopubli} 

\end{document}